\documentclass[journal = jpclcd, layout =onecolumn , manuscript = letter]{achemso}
\usepackage{amsmath}%author-numerical,%
\usepackage{amsfonts}
\usepackage{amssymb,bm}
\usepackage{graphicx}
\usepackage{physics}
\usepackage{upgreek}
\usepackage[outercaption]{sidecap}   
\usepackage{color}
\usepackage{hyperref}
\hypersetup{colorlinks=true,
	linkcolor=magenta,
	allcolors=magenta}
\usepackage{newtxtext}
\usepackage{newtxmath}
\usepackage[version=4]{mhchem}

\DeclareMathAlphabet{\pazocal}{OMS}{zplm}{m}{n}

\newcommand{\sys}{\ensuremath{\mathrm{s}}}
\newcommand{\nuc}{\ensuremath{\mathrm{n}}}
\newcommand{\el}{\ensuremath{\mathrm{e}}}

\newcommand{\sing}{\ensuremath{\mathrm{S}}}
\newcommand{\trip}{\ensuremath{\mathrm{T}}}

\renewcommand{\op}[1]{\ensuremath{\hat{#1}}}

\usepackage{booktabs}

	\title{Chirality Induced Spin Coherence in Electron Transfer Reactions}
	\author{Thomas P. Fay}
	\email{tom.patrick.fay@gmail.com}
	\affiliation{Department of Chemistry, University of Oxford, Physical and Theoretical Chemistry Laboratory, South Parks Road, Oxford, OX1 3QZ, UK}
\begin{document}

\begin{tocentry}
	\includegraphics{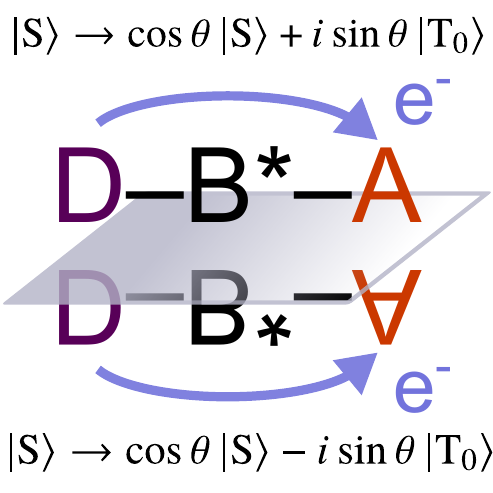}
\end{tocentry}	

\begin{abstract}
	Recently there has been much interest in the chirality induced spin selectivity effect, whereby electron spin polarisation, which is dependent on the molecular chirality, is produced in electrode-molecule electron transfer processes. Naturally, one might consider if a similar effect can be observed in simple molecular charge transfer reactions, for example in light-induced electron transfer from an electron donor to an electron acceptor. In this work we explore the effect of electron transfer on spins in chiral single radicals and chiral radical pairs using Nakajima-Zwanzig theory. In these cases chirality, in conjuction with spin-orbit coupling, does not lead to spin polarisation, but instead the electron transfer generates quantum coherence between spins states. In principle, this chirality induced spin coherence could manifest in a range of experiments, and in particular we demonstrate that the OOP-ESEEM pulse EPR experiment would be able to detect this effect in oriented radical pairs.
\end{abstract}

	\maketitle
	
%\section{Introduction}

Recently there has been significant progress in experiments\cite{Naaman2020,Torres-Cavanillas2020,Blumenschein2020,Rahman2020,Kulkarni2020,Metzger2020,Abendroth2019,Naaman2012} on and theoretical descriptions\cite{Gutierrez2012,Dalum2019,Michaeli2019} of the chirality induced spin selectivity (CISS) effect. Much of this has been motivated by the potential for this effect to be exploited in spintronic devices,\cite{Naaman2015} in which spins can be used to transfer and store information, with the potential for use in quantum computation and quantum information transfer.\cite{Naaman2015,Sun2014} The CISS effect involves the generation of electron spin polarisation by the transmission of an electron through a chiral environment, and this effect is generally explained by the presence of the spin-orbit interaction in these systems.\cite{Naaman2012,Gutierrez2012,Dalum2019} %, which couples the linear momentum of electrons travelling through a chiral medium to their spin, inducing a spin polarisation.\cite{Naaman2012,Gutierrez2012,Dalum2019} 
The majority of experimental and theoretical studies involve the transmission of electrons between electrodes and chiral molecules,\cite{Naaman2012,Naaman2015,Naaman2020} however there has been relatively little investigation into the role of chirality on the spin in molecular electron transfers, and so this is precisely what we explore here for reactions in involving radicals and radical pair formation.

It is instructive to first consider a charge transfer reaction in an $S=1/2$ doublet system,
\begin{align*}
\ce{{}^2[D^{$\bullet-$}A] <=>[$k_\mathrm{f}$][$k_\mathrm{b}$]{}^2[DA^{$\bullet-$}]}.
\end{align*}
A simple model for this is a system which can exist in one of two charge transfer (CT) states: $\ket{0} = \ket{\ce{D^{$\bullet -$}A}}$ and $\ket{1} = \ket{\ce{DA^{$\bullet -$}}}$.\cite{May2000,Nitzan2006,VanVoorhis2010,Fay2018} Associated with each of these diabatic states are two spin states $\ket{M_S = \pm 1/2}$, so overall there are four electronic states $\ket{j,M_S}$. For this system we can write the total Hamiltonian as\cite{Fay2018,May2000,Nitzan2006,VanVoorhis2010}
\begin{align}\label{tot-ham-eq}
	\op{H} = \op{H}_{0}\op{\Pi}_0 +  \op{H}_{1}\op{\Pi}_1 + \op{V}_{\mathrm{DC}} + \op{V}_{\mathrm{SOC}},
\end{align}
where $\op{\Pi}_j = \sum_{M_S=\pm1/2} \dyad{j,M_S}$ is a projection operator onto charge transfer state $j$, $\op{H}_j$ is the Hamiltonian for charge transfer state $j$, $\op{V}_{\mathrm{DC}}$ is the diabatic coupling Hamiltonian and $\op{V}_{\mathrm{SOC}}$ is the spin-orbit coupling Hamiltonian. The Hamiltonian for each charge transfer state $\op{H}_j$ can be divided into a nuclear part $\op{H}_{j\nuc} = \op{T}_\nuc + V_j(\op{\vb{Q}})$ which is a sum of nuclear kinetic energy and potential energy terms, and a spin part $\op{H}_{j\sys}$, and for simplicity we can assume the spin part is independent of the nuclear coordinates $\vb{Q}$.\cite{May2000,Nitzan2006}

The diabatic coupling term $\op{V}_{\mathrm{DC}}$ does not act on the spin of the system, instead it simply couples the two charge transfer states, and within the Condon approximation (coupling independent of $\vb{Q}$) this is\cite{Fay2018,May2000,Nitzan2006}
\begin{align}
	\op{V}_{\mathrm{DC}} = \Delta(\dyad{0}{1} + \dyad{1}{0})
\end{align}
where $\Delta$ is a real-valued constant. The spin-orbit coupling also couples the charge transfer states, but it also acts on the spin as follows\cite{Fedorov2003,Fedorov2005,Salem1972}
\begin{align}\label{Vsoc-radical-eq}
	\op{V}_{\mathrm{SOC}} = -i(\dyad{0}{1}\boldsymbol{\Lambda}\cdot \op{\vb{S}} - \boldsymbol{\Lambda}\cdot \op{\vb{S}}\dyad{1}{0}).
\end{align}
$\op{\vb{S}}$ here is the unitless electron spin operator and the vector coupling $\boldsymbol{\Lambda}$ is real valued for the spin-orbit coupling between bound electronic states (a detailed justification of this is given in the SI). Overall this means we can write the total charge transfer state coupling Hamiltonian $\op{V} =\op{V}_{\mathrm{DC}} + \op{V}_\mathrm{SOC}$ as
\begin{align}
	\op{V} = \Gamma\left(\op{U}^\dag \dyad{0}{1} + \dyad{1}{0}\op{U}\right),
\end{align}
where $\op{U}$ is a unitary spin operator given by $\op{U} = (\Delta + i \boldsymbol{\Lambda}\cdot \op{\vb{S}})/\Gamma$, and $\Gamma =\sqrt{\Delta^2 + |\boldsymbol{\Lambda}|^2/4}$. This unitary operator rotates the electron spin about the axis $\vb{n} =\boldsymbol{\Lambda}/|\Lambda|$ by an angle $2\theta$, with $\theta$ given by
\begin{align}\label{theta-eq}
\theta = \atan(|\boldsymbol{\Lambda}|/(2\Delta)).% CHECK
\end{align}

Coherences between the spin states of each charge transfer state will be long-lived because the potential energy surfaces for different spin states $V_{jM_S}(\vb{Q}) = V_j(\vb{Q})$ are identical, but this is not true for coherences between the different charge transfer states, and as such we describe the system with two spin density operators, one for each charge transfer state, $\op{\sigma}_{j\sys}(t)$ which are related to the full density operator $\op{\rho}(t)$ by\cite{Fay2018}
\begin{align}\label{sigmajs-def-eq}
	\op{\sigma}_{j\sys}(t) = \Tr_\nuc[\ev{\op{\rho}(t)}{j}].
\end{align}
Our aim is now to find a quantum master equation for these spin density operators. This is done by employing the Nakajima-Zwanzig equation\cite{Nakajima1958,Zwanzig1960} following the same steps as outlined in the derivation of radical pair spin master equations in Ref.~\citenum{Fay2018}, which we summarise in the SI. Assuming that the charge transfer coupling $\op{V}$ can be treated perturbatively, and that the population transfer is described accurately as an incoherent kinetic process, with rates independent of the spin interactions in $\op{H}_{j\sys}$, the resulting quantum master equations at second order in $\Gamma$ for $\op{\sigma}_{j\sys}(t)$ are,
\begin{align}
	\dv{t}\op{\sigma}_{0\sys}(t) &= -\frac{i}{\hbar}[\op{H}_{0\sys},\op{\sigma}_{1\sys}(t) ]-k_\mathrm{f} \op{\sigma}_{0\sys}(t) + k_\mathrm{b}\op{U}^\dag\op{\sigma}_{1\sys}\op{U} \\
	\dv{t}\op{\sigma}_{1\sys}(t) &= -\frac{i}{\hbar}[\op{H}_{1\sys},\op{\sigma}_{1\sys}(t) ]-k_\mathrm{b} \op{\sigma}_{1\sys}(t) + k_\mathrm{f}\op{U}\op{\sigma}_{0\sys}\op{U}^\dag,
\end{align}
where $[\cdot,\cdot]$ denotes the commutator. We see that as an electron is transferred, its spin is rotated about the vector $\boldsymbol{\Lambda}$, and as it is transferred back it is rotated about the vector $-\boldsymbol{\Lambda}$. 

Thus far we do not appear to have included the idea of chirality in any of the above discussion, so we shall now discuss how this manifests in these equations. Chirality of the system determines the relative sign of $\Delta$ and $\boldsymbol{\Lambda}$. On spatial reflection, the sign of the diabatic coupling $\Delta$ in unchanged, but the sign of the spin-orbit coupling vector $\boldsymbol{\Lambda}$ is changed because the spin-orbit coupling operator is an axial vector operator.\cite{Dalum2019} This means that for an initially spin polarised state of $\ket{0}$, $\op{\sigma}_{0s}(0) = \dyad{\alpha}$,\footnote{Here we use the standard notation for the electron spin states $\ket{\alpha} = \ket{M_S = +1/2}$ and $\ket{\beta} = \ket{M_S = -1/2}$.} this polarisation will be transformed into coherence between spin states $\ket{\alpha}$ and $\ket{\beta}$ in CT state $\ket{1}$ if $\boldsymbol{\Lambda}$ has a component perpendicular to the $z$ axis.\footnote{This rotation of the electron spin is reminiscent of the $\vec{C}$ term appearing in the theory of Dalum \& Hedeg{\aa}rd in Ref.~\citenum{Dalum2019}, derived for spin transport through a molecular junction from the non-equilibrium Green's function approach. A spin polarisation term analogous to the $\vec{D}$ term in in Ref.~\citenum{Dalum2019} does not appear in our theory for spin transport between bound electronic states because the initial and final electronic states have no linear or angular momentum, so their wavefunctions are real valued and the spin-orbit coupling vector is purely real, so no spin polarisation appears.} Interestingly this means that photo-initiated electron transfer in a spatially oriented chiral doublet system, initially at thermal equilibrium in a magnetic field, would yield $x/y$ magnetisation, and therefore a free-induction decay signal, \textit{without} a microwave pulse. For an achiral system there are equal contributions from configurations with $\pm\boldsymbol{\Lambda}$, %and thus the terms depending on $\boldsymbol{\Lambda}\Delta$ in the above expressions vanish, 
which means that spin polarisation perpendicular to the $\boldsymbol{\Lambda}$ axis is simply lost in the charge transfer process, and as such microwave pulse-free free-induction decay could not be observed.

%\subsection{The radical pair system}

We are now in a position to consider the effect of spin-orbit coupling and chirality on the charge transfer process between a singlet photoexcited precursor state $\ce{DA^*}$ and a charge separated radical pair state $\ce{D^{$\bullet +$}A^{$\bullet -$}}$, 
\begin{align*}
	\ce{{}^1[DA^*] <=>[$k_\mathrm{f}$][$k_\mathrm{b}$]{}^{$2S+1$}[D^{$\bullet+$}A^{$\bullet-$}]}.
\end{align*}
 Analogous to the above situation, a simple model for this includes the excited precursor diabatic state $\ket{0} = \ket{\ce{DA^*}}$ and a charge separated radical pair state $\ket{1} = \ket{\ce{D^{$\bullet +$}A^{$\bullet -$}}}$. We assume that the state $\ket{0}$ can only exist in a singlet spin state, so only $\ket{0,S=0,M_S=0}$ exists, but the radical pair state $\ket{1}$ can exist in both singlet and triplet states which lie very close in energy, therefore there are four (near-)degenerate $\ket{1,S,M_S}$ states which exist, with $S=0$ and $S=1$. The total Hamiltonian for this system has the same form as Eq.~\eqref{tot-ham-eq}, 
%\begin{align}
$	\op{H} = \op{H}_{0}\op{\Pi}_0 +  \op{H}_{1}\op{\Pi}_1 + \op{V}_{\mathrm{DC}} + \op{V}_{\mathrm{SOC}},$ 
%\end{align}
where again $\op{H}_j = \op{H}_{j\sys} + \op{H}_{j\nuc}$ is the Hamiltonian for state $j$ and $\op{\Pi}_j$ is a projection operator onto the $j$ states of the allowed spin multiplicity for that charge transfer state, i.e. $\op{\Pi}_0 = \dyad{0}\op{P}_\sing$ and $\op{\Pi}_1 = \dyad{1}(\op{P}_\sing + \op{P}_\trip)$, where $\op{P}_\sing$ and $\op{P}_\trip$ are singlet and triplet spin state projection operators. As before 
\begin{align}
\op{V}_{\mathrm{DC}} = \Delta (\op{P}_\sing\dyad{0}{1}+\dyad{1}{0}\op{P}_\sing)
\end{align}
is the diabatic coupling Hamiltonian, which now comes with a singlet projection operator $\op{P}_\sing$ because the diabatic coupling is spin conserving.\cite{Fay2018} The electron transfer can be regarded as a one electron transfer process, where only electron 1 transfers, and therefore the spin-orbit coupling term can be written as
\begin{align}\label{Vsoc-radicalpair-eq}
	\op{V}_{\mathrm{SOC}} = -i(\op{P}_\sing\dyad{0}{1}\boldsymbol{\Lambda}\cdot \op{\vb{S}}_1 - \boldsymbol{\Lambda}\cdot \op{\vb{S}}_1\dyad{1}{0}\op{P}_\sing).
\end{align}
where $\op{\vb{S}}_1$ is the spin operator for electron 1 (this is justified in more detail in the SI). This means the total charge transfer Hamiltonian can be written as
\begin{align}
	\op{V} = \Gamma (\op{P}_\sing\op{U}^\dag\dyad{0}{1}+\dyad{1}{0}\op{U}\op{P}_\sing)
\end{align}
where now $\op{U} = (\Delta + i \boldsymbol{\Lambda}\cdot \op{\vb{S}}_1)/\Gamma$ only acts on electron 1. 

From this we can obtain the quantum master equations for the spin density operators of the two charge transfer states, once again using the Nakajima-Zwanzig equation as in Ref.~\citenum{Fay2018}, which gives
\begin{align}
\begin{split}
	\dv{t}\op{\sigma}_{0\sys}(t) &= -\frac{i}{\hbar}[\op{H}_{0\sys},\op{\sigma}_{0\sys}(t) ]-k_\mathrm{f} \op{\sigma}_{0\sys}(t) 
	+k_\mathrm{b} \op{P}_\sing\op{U}^\dag\op{\sigma}_{1\sys}\op{U} \op{P}_\sing 
\end{split}
\\
\begin{split}
	\dv{t}\op{\sigma}_{1\sys}(t) &= -\frac{i}{\hbar}[\op{H}_{1\sys},\op{\sigma}_{1\sys}(t) ]-\left\{\frac{k_\mathrm{b}}{2}\op{U}\op{P}_\sing\op{U}^\dagger,\op{\sigma}_{1\sys}(t) \right\} 
	-\frac{i}{\hbar}\left[2\delta J\op{U}\op{P}_\sing\op{U}^\dagger,\op{\sigma}_{1\sys}(t) \right]+k_\mathrm{f}\op{U}\op{P}_\sing \op{\sigma}_{0\sys}(t)\op{P}_\sing\op{U}^\dag .
\end{split}
\end{align}
Here $\{\cdot,\cdot\}$ denotes the anti-commutator, and $2\delta J\op{U}\op{P}_\sing\op{U}^\dagger$ is an effective spin coupling term in the radical pair state (expressions for the master equation parameters $k_\mathrm{f}$, $k_\mathrm{b}$ and $\delta J$ are given in the SI). Now assuming the radical pair, state $\ket{1}$, is formed very rapidly and irreversibly from an excited singlet precursor $\ket{0}$, as is often the case for radical pairs formed by photoexcitation of a singlet ground state, then we can take $k_\mathrm{b} = 0$,
%and assuming that $k_\mathrm{f}$ is much larger than the frequencies associated with $\op{H}_{1s}$ and $\delta J$, then we can take 
and the radical pair spin density operator as being formed in the state
\begin{align}
	\op{\sigma}_{1\sys}(0) = \op{\sigma}_{\nuc\sys} \op{U}\op{P}_\sing\op{U}^\dag
\end{align}
where $\op{\sigma}_{\nuc\sys}$ is the initial nuclear spin state of the precursor state. The initial electron spin state can be written as 
%\begin{align}
$\op{U}\op{P}_\sing\op{U}^\dag = \dyad{\psi_0}{\psi_0},$ 
%\end{align}
where the initial spin state $\ket{\psi_0} = \op{U}\ket{\sing}$ is
\begin{align}
	\ket{\psi_0} = \cos\theta \ket{\sing} + i \sin\theta \ket{\trip_0 (\vb{n})}, 
\end{align}
in which $\ket{\sing}$ is an electron spin singlet state and $\ket{\trip_0 (\vb{n})}$ is a $\trip_0$ triplet state defined with respect to the axis $\vb{n} = \boldsymbol{\Lambda}/|\boldsymbol{\Lambda}|$, and $\theta$ is given by Eq.~\eqref{theta-eq}. Once again for the opposite enantiomer the sign of $\boldsymbol{\Lambda}$ changes. Denoting the different enantiomers $+$ and $-$, this means that $\vb{n}_+ = - \vb{n}_{-}$, or equivalently we can write the initial state as
\begin{align}
	\ket{\psi_0^{\pm}} = \cos\theta_+ \ket{\sing} \pm i \sin\theta_+ \ket{\trip_0 (\vb{n}_+)}.
\end{align}
This means that the phase of the initial coherence between singlet and triplet states is changed by the chirality in the radical pair. In an achiral system, there will be equal but opposite contributions to charge transfer from configurations with opposite signs of $\vb{n}$, and therefore there will be no initial coherence between singlet and triplet electron spin states. It is well known that spin-orbit coupled charge transport (SOCT) in radical pair reactions leads to the formation of triplet radical pairs,\cite{Salem1972,Levanon1978,Dance2008,Miura2010,Miura2010a,Colvin2012,Rehmat2020} but this shows that in chiral systems spin coherence between singlet and triplet states is also generated by SOCT. %By considering the spin-orbit coupling in detail, we see that in chiral systems spin-orbit coupled charge transport generates spin coherence between singlet and triplet states, in addition to chirality-independent triplet population. 

%\section{Detecting chirality induced spin coherence}

%In the next sections we will explore some methods for detecting this effect through standard EPR and MFE techniques. As a prerequisite for investigating any chirality induced spin effects in radical pairs, it is necessary to fix the orientation of the generated radical pairs in space. This could be achieved in a number of ways, for example by grafting the molecules consisting of covalently linked donors and acceptors to a surface in a fixed orientation. Assuming this orientation of the radical pairs can be realised, then the chirality induced spin coherence effect can be probed by EPR experiments or magnetic field effect experiments on the radical pair reaction.

%\subsection{OOP-ESEEM in oriented chiral radical pairs}
\begin{figure}[t]
	\includegraphics[width=0.45\textwidth]{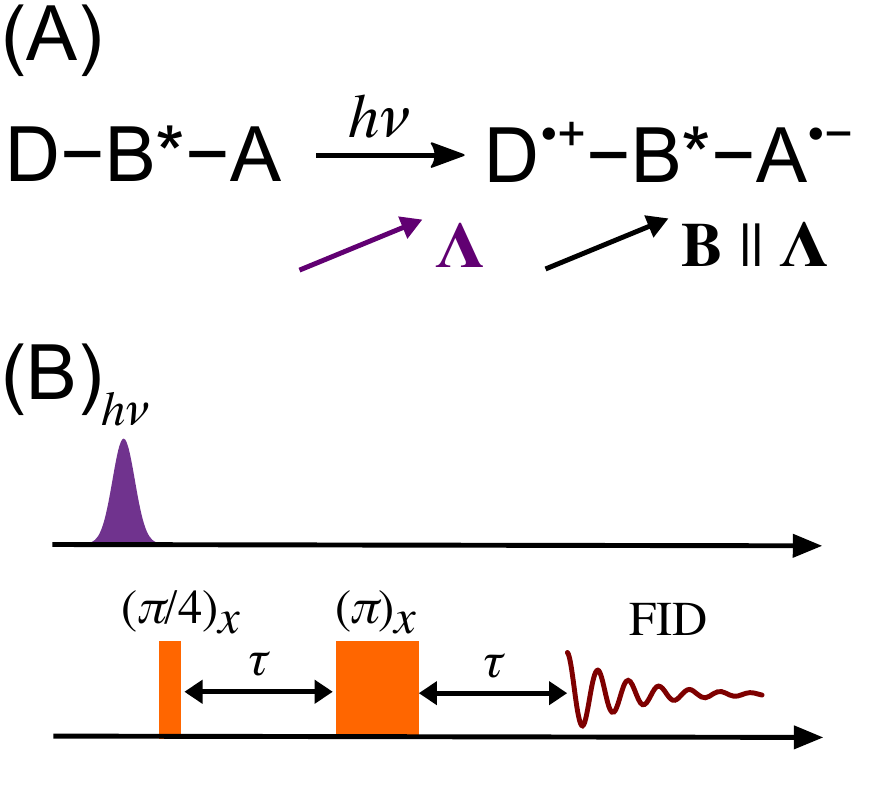}
	\caption{(A) A schematic representation radical pair formation in a magnetic field, with donor, \ce{D}, and acceptor, \ce{A}, covalently linked by a chiral bridge, \ce{B^*}. (B) A summary of the pulse sequence in the OOP-ESEEM experiment with $t_0 = 0$.}\label{exp-diag-fig}
\end{figure}
We will now consider how chirality induced spin coherence could be detected experimentally by EPR. Because the spin coherence generated is orientation dependent, any experiment probing this would have to be performed on radical pairs somehow fixed in a specific spatial orientation. %, which could be achieved by grafting the molecules consisting of donors and acceptors linked covalently by a chiral bridge to a surface. 
Here we consider the OOP-ESEEM (out of phase electron spin echo envelope modulation) EPR experiment, which has been developed as a technique for probing zero-quantum coherences in photo-generated radical pairs, with the applied field $\vb{B}$ parallel to $\boldsymbol{\Lambda}$ for oriented radical pairs as illustrated in Fig.~\ref{exp-diag-fig} (A).\cite{Salikhov1992,Tan1994,Dzuba1995,Hoff1998,Bittl2005} The standard OOP-ESEEM experiment consists a laser flash ($h\nu$), in which the radical pairs are generated, followed by a waiting time of $t_0$, followed by a non-selective $(\pi/4)_x$ microwave pulse, and another $(\pi)_x$ pulse after a time $\tau$, and after another waiting time $\tau$ the free-induction decay spin echo is recorded over times $t$. This $h\nu-t_0-(\pi/4)_x-\tau-(\pi)_x-\tau-\mathrm{FID}$ pulse sequence is illustrated schematically in Fig.~\ref{exp-diag-fig} (B). This pulse sequence gives access to information about the zero quantum coherences in the spin density operator,\cite{Hoff1998} which evolve in the waiting times $t_0$ and $\tau$. If we are interested in the initial state of the spin density operator, in particular the phases of the initial zero-quantum coherences, then we should minimise the time for the coherences to evolve on their own, as well as any decoherence processes, and therefore here we only consider the case of $t_0=0$, as shown in Fig.~\ref{exp-diag-fig} (B). 
%We also see that the initial state $\ket{\psi_0^\pm}$ can be equivalently written as $\ket{\psi_0^{\pm}} = (1/\sqrt{2})(e^{\pm i\theta_+}\ket{\alpha_1\beta_2} - e^{\mp i\theta_+}\ket{\beta_1\alpha_2} )$, which is equivalent to a time-evolved $\ket{S}$ state in the high-field limit, where the evolution time is positive or negative for different enantiomers. Since it is known the OOP-ESEEM signals are sensitive to the initial evolution time, 

Using the standard high-field radical pair spin Hamiltonian in the rotating frame,\cite{Hoff1998} and invoking the Schulten-Wolynes semiclassical approximation for the nuclear spins in the radical pair,\cite{Schulten1978} we can obtain expressions for the $x$-channel OOP-ESEEM signal $f_x(t) = \Tr_\sys[(\op{S}_{1x} + \op{S}_{2 x})\op{\sigma}_{1\sys}(t)]$ (details of this are given in the SI). In the weak coupling limit ($|\Omega_1-\Omega_2|\gg |J|,|d|$) we find the OOP-ESEEM FID signal to be,
%\begin{align}\label{weak-coup-fx-eq}
%	\begin{split}
%		&f_x(t) = -\frac{1}{2} \sin \left(\frac{\theta }{2}\right) \sin \left(\frac{1}{3} (d-3 J) (t+2 \tau )\right) \\
%		&\times\bigg(e^{-t^2/2\tau_1^2} \!\!\left(\sin \left(\frac{\theta }{2}\right) \cos \left(t \Omega _1\right)\!-\!\sqrt{2} \cos \left(\frac{\theta }{2}\right) \sin \left(t \Omega _1\right)\right)
%		+e^{-t^2/2\tau_2^2} \!\!\left(\sin \left(\frac{\theta }{2}\right) \cos \left(t \Omega _2\right)\!+\!\sqrt{2} \cos \left(\frac{\theta }{2}\right) \sin \left(t \Omega _2\right)\right)\bigg).
%	\end{split}
%\end{align}
\begin{align}\label{weak-coup-fx-eq}
	\begin{split}
		&f_x(t) = \frac{1}{2} \sin \left({\theta }\right) \sin \left(\left(J-\frac{d}{3}\right) (t+2 \tau )\right) \sum_{i=1,2}\bigg(e^{-t^2/(2\tau_i^2)} \!\!\left(\sin \left({\theta }\right) \cos \left(\Omega _i t \right)\!-\!\sqrt{2} \cos \left({\theta }\right) \sin \left(\Omega _i t\right)\right)
\bigg).
	\end{split}
\end{align}
$\Omega_i$ is the resonant frequency of electron spin $i$ (in the rotating frame), $J$ is the scalar coupling constant and $d$ is the effective dipolar coupling constant, and $\tau_{i}^{-2} = \frac{1}{3}\sum_{k=1}^{N_i} a_{i,k}^2 I_{i,k}(I_{i,k}+1)$, where $a_{i,k}$ and $I_{i,k}$ are the hyperfine coupling constants and spin quantum numbers for the hyperfine coupled nuclei in radical $i$. For simplicity we have ignored spin relaxation and radical pair recombination processes. Importantly here we see that changing the sign of $\theta$ changes the sign of certain terms in the $x$ channel FID signal after the OOP-ESEEM pulse sequence. This means the signal would be sensitive to the chirality of the molecule. As described in the SI, we can also calculate the OOP-ESEEM signal in the intermediate--strong coupling limit numerically.  Often in OOP-ESEEM experiments, the waiting time $\tau$ is varied, so we also consider the integrated FID echo signal as a function of $\tau$, $F_x(\tau) = \int_0^\infty f_x(t)\dd{t}$.
%\textbf{CHECK $\theta$ VS $2\theta$ IN MATHEMATICA SCRIPT!}

\begin{figure}[t]
	\includegraphics[width=0.45\textwidth]{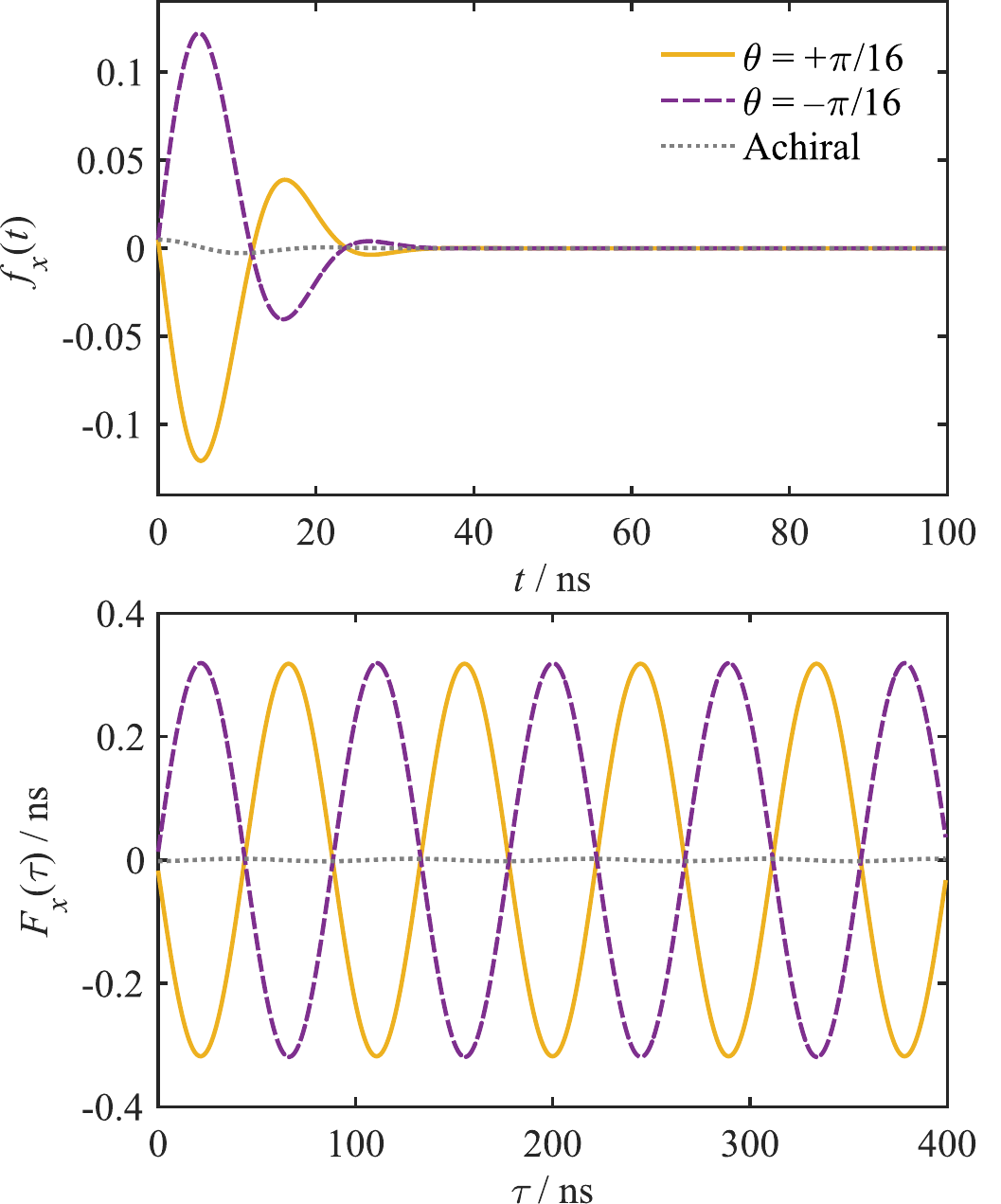}
	\caption{OOP-ESEEM signals for chiral radical pairs with $\Omega_1 = -\Omega_2 = 1.5\mathrm{\ mT}\ \gamma_\el$, $J-d/3 = 0.2\mathrm{\ mT}\ \gamma_\el$ and $\tau_1 = \tau_2 = 2.0 \mathrm{\ mT}^{-1}\ \gamma_\el^{-1}$, typical parameters for a real organic radical ion pair ($\gamma_\el = \el g_\el  / 2 m_\el$ is the modulus of the gyromagnetic ratio of the free electron spin). Achiral signals are an average of the $\theta = \pm \pi/16$ signals. Top panel: the FID echo for $\tau = 200\ \mathrm{ns}$ calculated using Eq.~\eqref{weak-coup-fx-eq}. Bottom panel: the integrated echo as a function of $\tau$ calculated using Eq.~(S.26). }\label{wc-eseem-fig}
\end{figure}
\begin{figure}[t]
	\includegraphics[width=0.45\textwidth]{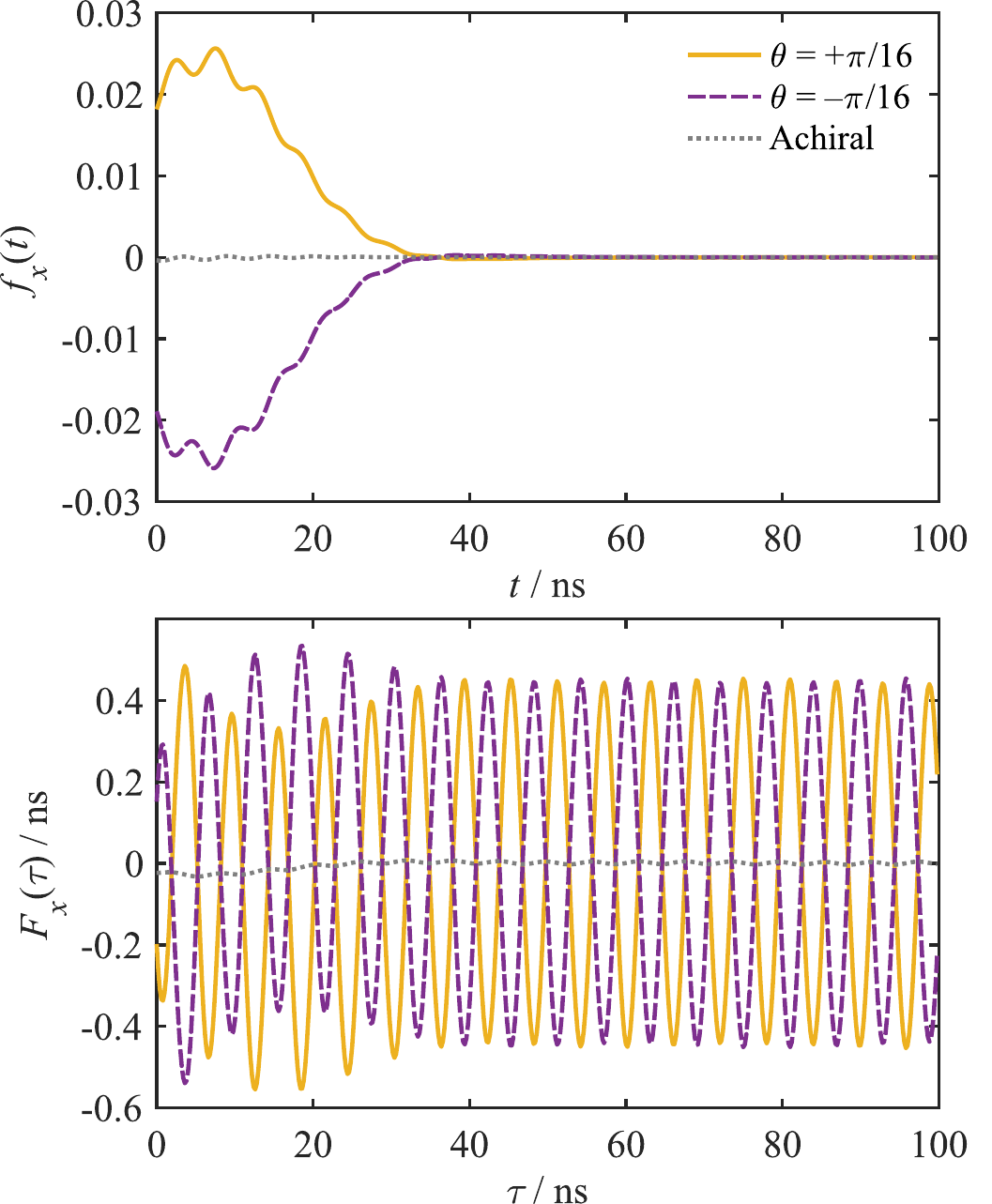}
	\caption{OOP-ESEEM signals for chiral radical pairs with $\Omega_1 = -\Omega_2 = 1.5\mathrm{\ mT}\ \gamma_\el$, $J-d/3 = 3\mathrm{\ mT}\ \gamma_\el$ and $\tau_1 = \tau_2 = 2.0 \mathrm{\ mT}^{-1}\ \gamma_\el^{-1}$. Achiral signals are an average of the $\theta = \pm \pi/16$ signals. Top panel: the FID echo for $\tau = 200\ \mathrm{ns}$ calculated using Eq.~(S.25). Bottom panel: the integrated echo as a function of $\tau$ calculated using Eq.~(S.25). }\label{sc-eseem-fig}
\end{figure}
As an example of this we plot the $t$ dependent OOP-ESEEM signal of two enantiomers of a chiral radical pair with $\theta = \pm \pi/16$ in Fig.~\ref{wc-eseem-fig}. This corresponds to an initial triplet fraction of $\sim\! 4\%$ which is typical of organic radical pairs in solution.\cite{Maeda2011a,Zhukov2020,Fay2019b} We see that the two OOP-ESEEM signals are very different even in the case of a modest SOCT contribution. The signals are not the exact negative of each other because the spin Hamiltonian is chirality independent but chirality does change the sign of the phase of coherences in the initial state, leading to a phase shift in the OOP-ESEEM signals. In Fig.~\ref{sc-eseem-fig} we show these signals for the same model in the intermediate coupling regime, calculated by sampling $10^6$ realisations of the hyperfine fields, where again we see that there is a phase shift between the $F_x(\tau)$ signals. 

% This of course may not be the only example of an experiment which can detect this effect; any experiment that allows one to measure the initial spin coherences would provide this information. In order to quantitatively interpret any real experiment, spin relaxation and recombination would need to be included in any model, but these results suggest that provided spin relaxation times and the radical pair lifetimes are in the order of 1 $\muup$s or more, then chirality induced spin coherence should be detectable in OOP-ESEEM experiments.

%\section{Conclusions}

In this work we have discussed how spin-orbit coupling can lead to chirality induced spin coherence in radicals and radical pair systems, an effect which has not before been explored in spin chemistry. In contrast to the case of electrode-molecule electron transfers, where chirality induces spin polarisation, chirality in intramolecular electron transfers generates specific coherent superpositions of spin states. This effect should manifest in certain EPR experiments and other experiments such as those that probe magnetic field effects (which we describe briefly in the SI). %We have also demonstrated that this effect could be detectable in oriented chiral radical pairs by standard EPR pulse experiments (and additionally in the SI we briefly describe how high field magnetic field effect experiments could also probe this).

Chirality induced spin coherence could be important in a range of systems. For example, it has been suggested that radical pair reactions in cryptochrome proteins form the basis of the magnetic compass sense of migratory birds,\cite{Ritz2000,Rodgers2009a,Wiltschko2019} and because chirality induced spin coherence is orientation dependent, this effect could hypothetically play a role in avian magnetoreception. Radical-based systems have also been proposed as potential molecular qubits,\cite{Rugg2017,Wu2018,Olshansky2019,Nelson2020,Wasielewski2020} and various quantum information theoretic ideas have been explored using these molecules.\cite{Rugg2019,Nelson2020,Wasielewski2020} Electron transport in chiral molecular systems could provide a way of further manipulating spin coherences in these systems. Other chiral open-shell molecular systems, aside from radicals and radical pairs, could potentially display chirality induced spin coherence effects. For example it has been demonstrated that spin-orbit coupling can play a role in singlet fission\cite{Basel2019} and triplet-triplet up-conversion,\cite{Ieuji2019} so it stands to reason that in chiral systems spin-orbit coupling could lead to interesting effects on these processes. Furthermore chirality induced spin coherence effects could manifest in the photophysics of open-shell chiral transition metal complexes with $S>0$. Overall we hope that this work will lay foundations for exploring chirality induced spin effects in a wide variety of molecular systems, beyond what has already been studied in electrode-molecule interfaces.

\section*{Acknowledgements}

I would like to thank Peter Hore and David Manolopoulos for their comments on this manuscript. I would also like to thank both Peter Hore and Jiate Luo for introducing me to the problem of spin effects in electron transfer in chiral molecules. I would like to acknowledge financial support from the Clarendon Scholarship from Oxford University, an E.A. Haigh Scholarship from Corpus Christi College, Oxford, the EPRSC Centre for Doctoral Training in Theory and Modelling in the Chemical Sciences, EPSRC Grant No. EP/L015722/1, and the Air Force Office of Scientific Research (Air Force Materiel Command, USAF award no. FA9550-14-1-0095).

\section*{Supporting Information}

Discussion and derivation of the spin-orbit coupling Hamiltonian in the diabatic representation used here, an outline of the second order Nakajima-Zwanzig master equation theory, expressions for OOP-ESEEM signals of chiral radical pairs in the intermediate-strong coupling limit, and a brief discussion of the effect of chirality induced spin coherence on quantum yields of radical pair reactions.

\bibliography{chiral-radical-pair-soct.bib}

\providecommand{\latin}[1]{#1}
\makeatletter
\providecommand{\doi}
  {\begingroup\let\do\@makeother\dospecials
  \catcode`\{=1 \catcode`\}=2 \doi@aux}
\providecommand{\doi@aux}[1]{\endgroup\texttt{#1}}
\makeatother
\providecommand*\mcitethebibliography{\thebibliography}
\csname @ifundefined\endcsname{endmcitethebibliography}
  {\let\endmcitethebibliography\endthebibliography}{}
\begin{mcitethebibliography}{49}
\providecommand*\natexlab[1]{#1}
\providecommand*\mciteSetBstSublistMode[1]{}
\providecommand*\mciteSetBstMaxWidthForm[2]{}
\providecommand*\mciteBstWouldAddEndPuncttrue
  {\def\EndOfBibitem{\unskip.}}
\providecommand*\mciteBstWouldAddEndPunctfalse
  {\let\EndOfBibitem\relax}
\providecommand*\mciteSetBstMidEndSepPunct[3]{}
\providecommand*\mciteSetBstSublistLabelBeginEnd[3]{}
\providecommand*\EndOfBibitem{}
\mciteSetBstSublistMode{f}
\mciteSetBstMaxWidthForm{subitem}{(\alph{mcitesubitemcount})}
\mciteSetBstSublistLabelBeginEnd
  {\mcitemaxwidthsubitemform\space}
  {\relax}
  {\relax}

\bibitem[Naaman \latin{et~al.}(2020)Naaman, Paltiel, and Waldeck]{Naaman2020}
Naaman,~R.; Paltiel,~Y.; Waldeck,~D.~H. {Chiral Induced Spin Selectivity Gives
  a New Twist on Spin-Control in Chemistry}. \emph{Acc. Chem. Res.}
  \textbf{2020}, \emph{53}, 2659--2667\relax
\mciteBstWouldAddEndPuncttrue
\mciteSetBstMidEndSepPunct{\mcitedefaultmidpunct}
{\mcitedefaultendpunct}{\mcitedefaultseppunct}\relax
\EndOfBibitem
\bibitem[Torres-Cavanillas \latin{et~al.}(2020)Torres-Cavanillas,
  Escorcia-Ariza, Brotons-Alc{\'{a}}zar, Sanchis-Gual, Mondal, Rosaleny,
  Gim{\'{e}}nez-Santamarina, Sessolo, Galbiati, Tatay, Gaita-Ari{\~{n}}o,
  Forment-Aliaga, and Cardona-Serra]{Torres-Cavanillas2020}
Torres-Cavanillas,~R.; Escorcia-Ariza,~G.; Brotons-Alc{\'{a}}zar,~I.;
  Sanchis-Gual,~R.; Mondal,~P.~C.; Rosaleny,~L.~E.;
  Gim{\'{e}}nez-Santamarina,~S.; Sessolo,~M.; Galbiati,~M.; Tatay,~S.
  \latin{et~al.}  {Reinforced Room-Temperature Spin Filtering in Chiral
  Paramagnetic Metallopeptides}. \emph{J. Am. Chem. Soc.} \textbf{2020},
  \emph{142}, 17572--17580\relax
\mciteBstWouldAddEndPuncttrue
\mciteSetBstMidEndSepPunct{\mcitedefaultmidpunct}
{\mcitedefaultendpunct}{\mcitedefaultseppunct}\relax
\EndOfBibitem
\bibitem[Blumenschein \latin{et~al.}(2020)Blumenschein, Tamski, Roussel,
  Smolinsky, Tassinari, Naaman, and Ansermet]{Blumenschein2020}
Blumenschein,~F.; Tamski,~M.; Roussel,~C.; Smolinsky,~E. Z.~B.; Tassinari,~F.;
  Naaman,~R.; Ansermet,~J.-P. {Spin-dependent charge transfer at chiral
  electrodes probed by magnetic resonance}. \emph{Phys. Chem. Chem. Phys.}
  \textbf{2020}, \emph{22}, 997--1002\relax
\mciteBstWouldAddEndPuncttrue
\mciteSetBstMidEndSepPunct{\mcitedefaultmidpunct}
{\mcitedefaultendpunct}{\mcitedefaultseppunct}\relax
\EndOfBibitem
\bibitem[Rahman \latin{et~al.}(2020)Rahman, Firouzeh, Mujica, and
  Pramanik]{Rahman2020}
Rahman,~M.~W.; Firouzeh,~S.; Mujica,~V.; Pramanik,~S. {Carrier Transport
  Engineering in Carbon Nanotubes by Chirality-Induced Spin Polarization}.
  \emph{ACS Nano} \textbf{2020}, \emph{14}, 3389--3396\relax
\mciteBstWouldAddEndPuncttrue
\mciteSetBstMidEndSepPunct{\mcitedefaultmidpunct}
{\mcitedefaultendpunct}{\mcitedefaultseppunct}\relax
\EndOfBibitem
\bibitem[Kulkarni \latin{et~al.}(2020)Kulkarni, Mondal, Das, Grinbom,
  Tassinari, Mabesoone, Meijer, and Naaman]{Kulkarni2020}
Kulkarni,~C.; Mondal,~A.~K.; Das,~T.~K.; Grinbom,~G.; Tassinari,~F.;
  Mabesoone,~M. F.~J.; Meijer,~E.~W.; Naaman,~R. {Highly Efficient and Tunable
  Filtering of Electrons' Spin by Supramolecular Chirality of Nanofiber‐Based
  Materials}. \emph{Adv. Mater.} \textbf{2020}, \emph{32}, 1904965\relax
\mciteBstWouldAddEndPuncttrue
\mciteSetBstMidEndSepPunct{\mcitedefaultmidpunct}
{\mcitedefaultendpunct}{\mcitedefaultseppunct}\relax
\EndOfBibitem
\bibitem[Metzger \latin{et~al.}(2020)Metzger, Mishra, Bloom, Goren, Neubauer,
  Shmul, Wei, Yochelis, Tassinari, Fontanesi, Waldeck, Paltiel, and
  Naaman]{Metzger2020}
Metzger,~T.~S.; Mishra,~S.; Bloom,~B.~P.; Goren,~N.; Neubauer,~A.; Shmul,~G.;
  Wei,~J.; Yochelis,~S.; Tassinari,~F.; Fontanesi,~C. \latin{et~al.}  {The
  Electron Spin as a Chiral Reagent}. \emph{Angew. Chemie - Int. Ed.}
  \textbf{2020}, \emph{59}, 1653--1658\relax
\mciteBstWouldAddEndPuncttrue
\mciteSetBstMidEndSepPunct{\mcitedefaultmidpunct}
{\mcitedefaultendpunct}{\mcitedefaultseppunct}\relax
\EndOfBibitem
\bibitem[Abendroth \latin{et~al.}(2019)Abendroth, Stemer, Bloom, Roy, Naaman,
  Waldeck, Weiss, and Mondal]{Abendroth2019}
Abendroth,~J.~M.; Stemer,~D.~M.; Bloom,~B.~P.; Roy,~P.; Naaman,~R.;
  Waldeck,~D.~H.; Weiss,~P.~S.; Mondal,~P.~C. {Spin Selectivity in Photoinduced
  Charge-Transfer Mediated by Chiral Molecules}. \emph{ACS Nano} \textbf{2019},
  \emph{13}, 4928--4946\relax
\mciteBstWouldAddEndPuncttrue
\mciteSetBstMidEndSepPunct{\mcitedefaultmidpunct}
{\mcitedefaultendpunct}{\mcitedefaultseppunct}\relax
\EndOfBibitem
\bibitem[Naaman and Waldeck(2012)Naaman, and Waldeck]{Naaman2012}
Naaman,~R.; Waldeck,~D.~H. {Chiral-Induced Spin Selectivity Effect}. \emph{J.
  Phys. Chem. Lett.} \textbf{2012}, \emph{3}, 2178--2187\relax
\mciteBstWouldAddEndPuncttrue
\mciteSetBstMidEndSepPunct{\mcitedefaultmidpunct}
{\mcitedefaultendpunct}{\mcitedefaultseppunct}\relax
\EndOfBibitem
\bibitem[Gutierrez \latin{et~al.}(2012)Gutierrez, D{\'{i}}az, Naaman, and
  Cuniberti]{Gutierrez2012}
Gutierrez,~R.; D{\'{i}}az,~E.; Naaman,~R.; Cuniberti,~G. {Spin-selective
  transport through helical molecular systems}. \emph{Phys. Rev. B}
  \textbf{2012}, \emph{85}, 081404\relax
\mciteBstWouldAddEndPuncttrue
\mciteSetBstMidEndSepPunct{\mcitedefaultmidpunct}
{\mcitedefaultendpunct}{\mcitedefaultseppunct}\relax
\EndOfBibitem
\bibitem[Dalum and Hedeg{\aa}rd(2019)Dalum, and Hedeg{\aa}rd]{Dalum2019}
Dalum,~S.; Hedeg{\aa}rd,~P. {Theory of Chiral Induced Spin Selectivity}.
  \emph{Nano Lett.} \textbf{2019}, \emph{19}, 5253--5259\relax
\mciteBstWouldAddEndPuncttrue
\mciteSetBstMidEndSepPunct{\mcitedefaultmidpunct}
{\mcitedefaultendpunct}{\mcitedefaultseppunct}\relax
\EndOfBibitem
\bibitem[Michaeli and Naaman(2019)Michaeli, and Naaman]{Michaeli2019}
Michaeli,~K.; Naaman,~R. {Origin of Spin-Dependent Tunneling Through Chiral
  Molecules}. \emph{J. Phys. Chem. C} \textbf{2019}, \emph{123},
  17043--17048\relax
\mciteBstWouldAddEndPuncttrue
\mciteSetBstMidEndSepPunct{\mcitedefaultmidpunct}
{\mcitedefaultendpunct}{\mcitedefaultseppunct}\relax
\EndOfBibitem
\bibitem[Naaman and Waldeck(2015)Naaman, and Waldeck]{Naaman2015}
Naaman,~R.; Waldeck,~D.~H. {Spintronics and Chirality: Spin Selectivity in
  Electron Transport Through Chiral Molecules}. \emph{Annu. Rev. Phys. Chem.}
  \textbf{2015}, \emph{66}, 263--281\relax
\mciteBstWouldAddEndPuncttrue
\mciteSetBstMidEndSepPunct{\mcitedefaultmidpunct}
{\mcitedefaultendpunct}{\mcitedefaultseppunct}\relax
\EndOfBibitem
\bibitem[Sun \latin{et~al.}(2014)Sun, Ehrenfreund, and {Valy Vardeny}]{Sun2014}
Sun,~D.; Ehrenfreund,~E.; {Valy Vardeny},~Z. {The first decade of organic
  spintronics research}. \emph{Chem. Commun.} \textbf{2014}, \emph{50},
  1781--1793\relax
\mciteBstWouldAddEndPuncttrue
\mciteSetBstMidEndSepPunct{\mcitedefaultmidpunct}
{\mcitedefaultendpunct}{\mcitedefaultseppunct}\relax
\EndOfBibitem
\bibitem[May and K\"uhn(2000)May, and K\"uhn]{May2000}
May,~V.; K\"uhn,~O. \emph{{Charge and Energy Transfer Dynamics in Molecular
  Systems}}; Wiley-VCH Verlag GmbH {\&} Co. KGaA, 2000\relax
\mciteBstWouldAddEndPuncttrue
\mciteSetBstMidEndSepPunct{\mcitedefaultmidpunct}
{\mcitedefaultendpunct}{\mcitedefaultseppunct}\relax
\EndOfBibitem
\bibitem[Nitzan(2006)]{Nitzan2006}
Nitzan,~A. \emph{{Chemical Dynamics in Condensed Phases}}; Oxford University
  Press, 2006\relax
\mciteBstWouldAddEndPuncttrue
\mciteSetBstMidEndSepPunct{\mcitedefaultmidpunct}
{\mcitedefaultendpunct}{\mcitedefaultseppunct}\relax
\EndOfBibitem
\bibitem[{Van Voorhis} \latin{et~al.}(2010){Van Voorhis}, Kowalczyk, Kaduk,
  Wang, Cheng, and Wu]{VanVoorhis2010}
{Van Voorhis},~T.; Kowalczyk,~T.; Kaduk,~B.; Wang,~L.-P.; Cheng,~C.-L.; Wu,~Q.
  {The Diabatic Picture of Electron Transfer, Reaction Barriers, and Molecular
  Dynamics}. \emph{Annu. Rev. Phys. Chem.} \textbf{2010}, \emph{61},
  149--170\relax
\mciteBstWouldAddEndPuncttrue
\mciteSetBstMidEndSepPunct{\mcitedefaultmidpunct}
{\mcitedefaultendpunct}{\mcitedefaultseppunct}\relax
\EndOfBibitem
\bibitem[Fay \latin{et~al.}(2018)Fay, Lindoy, and Manolopoulos]{Fay2018}
Fay,~T.~P.; Lindoy,~L.~P.; Manolopoulos,~D.~E. {Spin-selective electron
  transfer reactions of radical pairs: Beyond the Haberkorn master equation}.
  \emph{J. Chem. Phys.} \textbf{2018}, \emph{149}, 064107\relax
\mciteBstWouldAddEndPuncttrue
\mciteSetBstMidEndSepPunct{\mcitedefaultmidpunct}
{\mcitedefaultendpunct}{\mcitedefaultseppunct}\relax
\EndOfBibitem
\bibitem[Fedorov \latin{et~al.}(2003)Fedorov, Koseki, Schmidt, and
  Gordon]{Fedorov2003}
Fedorov,~D.~G.; Koseki,~S.; Schmidt,~M.~W.; Gordon,~M.~S. {Spin-orbit coupling
  in molecules: Chemistry beyond the adiabatic approximation}. \emph{Int. Rev.
  Phys. Chem.} \textbf{2003}, \emph{22}, 551--592\relax
\mciteBstWouldAddEndPuncttrue
\mciteSetBstMidEndSepPunct{\mcitedefaultmidpunct}
{\mcitedefaultendpunct}{\mcitedefaultseppunct}\relax
\EndOfBibitem
\bibitem[Fedorov and Gordon(2000)Fedorov, and Gordon]{Fedorov2005}
Fedorov,~D.~G.; Gordon,~M.~S. {A study of the relative importance of one and
  two-electron contributions to spin–orbit coupling}. \emph{J. Chem. Phys.}
  \textbf{2000}, \emph{112}, 5611--5623\relax
\mciteBstWouldAddEndPuncttrue
\mciteSetBstMidEndSepPunct{\mcitedefaultmidpunct}
{\mcitedefaultendpunct}{\mcitedefaultseppunct}\relax
\EndOfBibitem
\bibitem[Salem and Rowland(1972)Salem, and Rowland]{Salem1972}
Salem,~L.; Rowland,~C. {The Electronic Properties of Diradicals}. \emph{Angew.
  Chemie Int. Ed. English} \textbf{1972}, \emph{11}, 92--111\relax
\mciteBstWouldAddEndPuncttrue
\mciteSetBstMidEndSepPunct{\mcitedefaultmidpunct}
{\mcitedefaultendpunct}{\mcitedefaultseppunct}\relax
\EndOfBibitem
\bibitem[Nakajima(1958)]{Nakajima1958}
Nakajima,~S. {On Quantum Theory of Transport Phenomena}. \emph{Prog. Theor.
  Phys.} \textbf{1958}, \emph{20}, 948--959\relax
\mciteBstWouldAddEndPuncttrue
\mciteSetBstMidEndSepPunct{\mcitedefaultmidpunct}
{\mcitedefaultendpunct}{\mcitedefaultseppunct}\relax
\EndOfBibitem
\bibitem[Zwanzig(1960)]{Zwanzig1960}
Zwanzig,~R. {Ensemble method in the theory of irreversibility}. \emph{J. Chem.
  Phys.} \textbf{1960}, \emph{33}, 1338--1341\relax
\mciteBstWouldAddEndPuncttrue
\mciteSetBstMidEndSepPunct{\mcitedefaultmidpunct}
{\mcitedefaultendpunct}{\mcitedefaultseppunct}\relax
\EndOfBibitem
\bibitem[Levanon and Norris(1978)Levanon, and Norris]{Levanon1978}
Levanon,~H.; Norris,~J.~R. {The photoexcited triplet state and photosynthesis}.
  \emph{Chem. Rev.} \textbf{1978}, \emph{78}, 185--198\relax
\mciteBstWouldAddEndPuncttrue
\mciteSetBstMidEndSepPunct{\mcitedefaultmidpunct}
{\mcitedefaultendpunct}{\mcitedefaultseppunct}\relax
\EndOfBibitem
\bibitem[Dance \latin{et~al.}(2008)Dance, Mickley, Wilson, Ricks, Scott,
  Ratner, and Wasielewski]{Dance2008}
Dance,~Z. E.~X.; Mickley,~S.~M.; Wilson,~T.~M.; Ricks,~A.~B.; Scott,~A.~M.;
  Ratner,~M.~A.; Wasielewski,~M.~R. {Intersystem Crossing Mediated by
  Photoinduced Intramolecular Charge Transfer: Julolidine-Anthracene Molecules
  with Perpendicular $\pi$ Systems}. \emph{J. Phys. Chem. A} \textbf{2008},
  \emph{112}, 4194--4201\relax
\mciteBstWouldAddEndPuncttrue
\mciteSetBstMidEndSepPunct{\mcitedefaultmidpunct}
{\mcitedefaultendpunct}{\mcitedefaultseppunct}\relax
\EndOfBibitem
\bibitem[Miura \latin{et~al.}(2010)Miura, Scott, and Wasielewski]{Miura2010}
Miura,~T.; Scott,~A.~M.; Wasielewski,~M.~R. {Electron spin dynamics as a
  controlling factor for spin-selective charge recombination in donor - Bridge
  - Acceptor molecules}. \emph{J. Phys. Chem. C} \textbf{2010}, \emph{114},
  20370--20379\relax
\mciteBstWouldAddEndPuncttrue
\mciteSetBstMidEndSepPunct{\mcitedefaultmidpunct}
{\mcitedefaultendpunct}{\mcitedefaultseppunct}\relax
\EndOfBibitem
\bibitem[Miura \latin{et~al.}(2010)Miura, Carmieli, and
  Wasielewski]{Miura2010a}
Miura,~T.; Carmieli,~R.; Wasielewski,~M.~R. {Time-resolved EPR studies of
  charge recombination and triplet-state formation within donor-bridge-acceptor
  molecules having wire-like oligofluorene bridges}. \emph{J. Phys. Chem. A}
  \textbf{2010}, \emph{114}, 5769--5778\relax
\mciteBstWouldAddEndPuncttrue
\mciteSetBstMidEndSepPunct{\mcitedefaultmidpunct}
{\mcitedefaultendpunct}{\mcitedefaultseppunct}\relax
\EndOfBibitem
\bibitem[Colvin \latin{et~al.}(2012)Colvin, Ricks, Scott, Co, and
  Wasielewski]{Colvin2012}
Colvin,~M.~T.; Ricks,~A.~B.; Scott,~A.~M.; Co,~D.~T.; Wasielewski,~M.~R.
  {Intersystem crossing involving strongly spin exchange-coupled radical ion
  pairs in donor-bridge-acceptor molecules}. \emph{J. Phys. Chem. A}
  \textbf{2012}, \emph{116}, 1923--1930\relax
\mciteBstWouldAddEndPuncttrue
\mciteSetBstMidEndSepPunct{\mcitedefaultmidpunct}
{\mcitedefaultendpunct}{\mcitedefaultseppunct}\relax
\EndOfBibitem
\bibitem[Rehmat \latin{et~al.}(2020)Rehmat, Toffoletti, Mahmood, Zhang, Zhao,
  and Barbon]{Rehmat2020}
Rehmat,~N.; Toffoletti,~A.; Mahmood,~Z.; Zhang,~X.; Zhao,~J.; Barbon,~A.
  {Carbazole-perylenebisimide electron donor/acceptor dyads showing efficient
  spin orbit charge transfer intersystem crossing (SOCT-ISC) and photo-driven
  intermolecular electron transfer}. \emph{J. Mater. Chem. C} \textbf{2020},
  \emph{8}, 4701--4712\relax
\mciteBstWouldAddEndPuncttrue
\mciteSetBstMidEndSepPunct{\mcitedefaultmidpunct}
{\mcitedefaultendpunct}{\mcitedefaultseppunct}\relax
\EndOfBibitem
\bibitem[Salikhov \latin{et~al.}(1992)Salikhov, Kandrashkin, and
  Salikhov]{Salikhov1992}
Salikhov,~K.~M.; Kandrashkin,~Y.~E.; Salikhov,~A.~K. {Peculiarities of free
  induction and primary spin echo signals for spin-correlated radical pairs}.
  \emph{Appl. Magn. Reson.} \textbf{1992}, \emph{3}, 199--216\relax
\mciteBstWouldAddEndPuncttrue
\mciteSetBstMidEndSepPunct{\mcitedefaultmidpunct}
{\mcitedefaultendpunct}{\mcitedefaultseppunct}\relax
\EndOfBibitem
\bibitem[Tan \latin{et~al.}(1994)Tan, Thurnauer, and Norris]{Tan1994}
Tan,~J.; Thurnauer,~M.~C.; Norris,~J.~R. {Electron spin echo envelope
  modulation due to exchange and dipolar interactions in a spin-correlated
  radical pair}. \emph{Chem. Phys. Lett.} \textbf{1994}, \emph{219},
  283--290\relax
\mciteBstWouldAddEndPuncttrue
\mciteSetBstMidEndSepPunct{\mcitedefaultmidpunct}
{\mcitedefaultendpunct}{\mcitedefaultseppunct}\relax
\EndOfBibitem
\bibitem[Dzuba \latin{et~al.}(1995)Dzuba, Gast, and Hoff]{Dzuba1995}
Dzuba,~S.~A.; Gast,~P.; Hoff,~A.~J. {ESEEM study of spin-spin interactions in
  spin-polarised P+QA- pairs in the photosynthetic purple bacterium Rhodobacter
  sphaeroides R26}. \emph{Chem. Phys. Lett.} \textbf{1995}, \emph{236},
  595--602\relax
\mciteBstWouldAddEndPuncttrue
\mciteSetBstMidEndSepPunct{\mcitedefaultmidpunct}
{\mcitedefaultendpunct}{\mcitedefaultseppunct}\relax
\EndOfBibitem
\bibitem[Hoff \latin{et~al.}(1998)Hoff, Gast, Dzuba, Timmel, Fursman, and
  Hore]{Hoff1998}
Hoff,~A.~J.; Gast,~P.; Dzuba,~S.~A.; Timmel,~C.~R.; Fursman,~C.~E.; Hore,~P.
  {The nuts and bolts of distance determination and zero- and double-quantum
  coherence in photoinduced radical pairs}. \emph{Spectrochim. Acta Part A Mol.
  Biomol. Spectrosc.} \textbf{1998}, \emph{54}, 2283--2293\relax
\mciteBstWouldAddEndPuncttrue
\mciteSetBstMidEndSepPunct{\mcitedefaultmidpunct}
{\mcitedefaultendpunct}{\mcitedefaultseppunct}\relax
\EndOfBibitem
\bibitem[Bittl and Weber(2005)Bittl, and Weber]{Bittl2005}
Bittl,~R.; Weber,~S. {Transient radical pairs studied by time-resolved EPR}.
  \emph{Biochim. Biophys. Acta - Bioenerg.} \textbf{2005}, \emph{1707},
  117--126\relax
\mciteBstWouldAddEndPuncttrue
\mciteSetBstMidEndSepPunct{\mcitedefaultmidpunct}
{\mcitedefaultendpunct}{\mcitedefaultseppunct}\relax
\EndOfBibitem
\bibitem[Schulten and Wolynes(1978)Schulten, and Wolynes]{Schulten1978}
Schulten,~K.; Wolynes,~P.~G. {Semiclassical description of electron spin motion
  in radicals including the effect of electron hopping}. \emph{J. Chem. Phys.}
  \textbf{1978}, \emph{68}, 3292--3297\relax
\mciteBstWouldAddEndPuncttrue
\mciteSetBstMidEndSepPunct{\mcitedefaultmidpunct}
{\mcitedefaultendpunct}{\mcitedefaultseppunct}\relax
\EndOfBibitem
\bibitem[Maeda \latin{et~al.}(2011)Maeda, Wedge, Storey, Henbest, Liddell,
  Kodis, Gust, Hore, and Timmel]{Maeda2011a}
Maeda,~K.; Wedge,~C.~J.; Storey,~J.~G.; Henbest,~K.~B.; Liddell,~P.~A.;
  Kodis,~G.; Gust,~D.; Hore,~P.~J.; Timmel,~C.~R. {Spin-selective recombination
  kinetics of a model chemical magnetoreceptor}. \emph{Chem. Commun.}
  \textbf{2011}, \emph{47}, 6563--6565\relax
\mciteBstWouldAddEndPuncttrue
\mciteSetBstMidEndSepPunct{\mcitedefaultmidpunct}
{\mcitedefaultendpunct}{\mcitedefaultseppunct}\relax
\EndOfBibitem
\bibitem[Zhukov \latin{et~al.}(2020)Zhukov, Fishman, Kiryutin, Lukzen, Panov,
  Steiner, Vieth, Sch{\"{a}}fer, Lambert, and Yurkovskaya]{Zhukov2020}
Zhukov,~I.; Fishman,~N.; Kiryutin,~A.; Lukzen,~N.; Panov,~M.; Steiner,~U.;
  Vieth,~H.-M.; Sch{\"{a}}fer,~J.; Lambert,~C.; Yurkovskaya,~A. {Positive
  electronic exchange interaction and predominance of minor triplet channel in
  CIDNP formation in short lived charge separated states of D-X-A dyads}.
  \emph{J. Chem. Phys.} \textbf{2020}, \emph{152}, 014203\relax
\mciteBstWouldAddEndPuncttrue
\mciteSetBstMidEndSepPunct{\mcitedefaultmidpunct}
{\mcitedefaultendpunct}{\mcitedefaultseppunct}\relax
\EndOfBibitem
\bibitem[Fay \latin{et~al.}(2019)Fay, Lindoy, and Manolopoulos]{Fay2019b}
Fay,~T.~P.; Lindoy,~L.~P.; Manolopoulos,~D.~E. {Electron spin relaxation in
  radical pairs: Beyond the Redfield approximation}. \emph{J. Chem. Phys.}
  \textbf{2019}, \emph{151}, 154117\relax
\mciteBstWouldAddEndPuncttrue
\mciteSetBstMidEndSepPunct{\mcitedefaultmidpunct}
{\mcitedefaultendpunct}{\mcitedefaultseppunct}\relax
\EndOfBibitem
\bibitem[Ritz \latin{et~al.}(2000)Ritz, Adem, and Schulten]{Ritz2000}
Ritz,~T.; Adem,~S.; Schulten,~K. {A Model for Photoreceptor-Based
  Magnetoreception in Birds}. \emph{Biophys. J.} \textbf{2000}, \emph{78},
  707--718\relax
\mciteBstWouldAddEndPuncttrue
\mciteSetBstMidEndSepPunct{\mcitedefaultmidpunct}
{\mcitedefaultendpunct}{\mcitedefaultseppunct}\relax
\EndOfBibitem
\bibitem[Rodgers and Hore(2009)Rodgers, and Hore]{Rodgers2009a}
Rodgers,~C.~T.; Hore,~P.~J. {Chemical magnetoreception in birds: The radical
  pair mechanism}. \emph{Proc. Natl. Acad. Sci.} \textbf{2009}, \emph{106},
  353--360\relax
\mciteBstWouldAddEndPuncttrue
\mciteSetBstMidEndSepPunct{\mcitedefaultmidpunct}
{\mcitedefaultendpunct}{\mcitedefaultseppunct}\relax
\EndOfBibitem
\bibitem[Wiltschko and Wiltschko(2019)Wiltschko, and Wiltschko]{Wiltschko2019}
Wiltschko,~R.; Wiltschko,~W. {Magnetoreception in birds}. \emph{J. R. Soc.
  Interface} \textbf{2019}, \emph{16}, 20190295\relax
\mciteBstWouldAddEndPuncttrue
\mciteSetBstMidEndSepPunct{\mcitedefaultmidpunct}
{\mcitedefaultendpunct}{\mcitedefaultseppunct}\relax
\EndOfBibitem
\bibitem[Rugg \latin{et~al.}(2017)Rugg, Phelan, Horwitz, Young, Krzyaniak,
  Ratner, and Wasielewski]{Rugg2017}
Rugg,~B.~K.; Phelan,~B.~T.; Horwitz,~N.~E.; Young,~R.~M.; Krzyaniak,~M.~D.;
  Ratner,~M.~A.; Wasielewski,~M.~R. {Spin-Selective Photoreduction of a Stable
  Radical within a Covalent Donor-Acceptor-Radical Triad}. \emph{J. Am. Chem.
  Soc.} \textbf{2017}, \emph{139}, 15660--15663\relax
\mciteBstWouldAddEndPuncttrue
\mciteSetBstMidEndSepPunct{\mcitedefaultmidpunct}
{\mcitedefaultendpunct}{\mcitedefaultseppunct}\relax
\EndOfBibitem
\bibitem[Wu \latin{et~al.}(2018)Wu, Zhou, Nelson, Young, Krzyaniak, and
  Wasielewski]{Wu2018}
Wu,~Y.; Zhou,~J.; Nelson,~J.~N.; Young,~R.~M.; Krzyaniak,~M.~D.;
  Wasielewski,~M.~R. {Covalent Radical Pairs as Spin Qubits: Influence of Rapid
  Electron Motion between Two Equivalent Sites on Spin Coherence}. \emph{J. Am.
  Chem. Soc.} \textbf{2018}, \emph{140}, 13011--13021\relax
\mciteBstWouldAddEndPuncttrue
\mciteSetBstMidEndSepPunct{\mcitedefaultmidpunct}
{\mcitedefaultendpunct}{\mcitedefaultseppunct}\relax
\EndOfBibitem
\bibitem[Olshansky \latin{et~al.}(2019)Olshansky, Krzyaniak, Young, and
  Wasielewski]{Olshansky2019}
Olshansky,~J.~H.; Krzyaniak,~M.~D.; Young,~R.~M.; Wasielewski,~M.~R.
  {Photogenerated Spin-Entangled Qubit (Radical) Pairs in DNA Hairpins:
  Observation of Spin Delocalization and Coherence}. \emph{J. Am. Chem. Soc.}
  \textbf{2019}, \emph{141}, 2152--2160\relax
\mciteBstWouldAddEndPuncttrue
\mciteSetBstMidEndSepPunct{\mcitedefaultmidpunct}
{\mcitedefaultendpunct}{\mcitedefaultseppunct}\relax
\EndOfBibitem
\bibitem[Nelson \latin{et~al.}(2020)Nelson, Zhang, Zhou, Rugg, Krzyaniak, and
  Wasielewski]{Nelson2020}
Nelson,~J.~N.; Zhang,~J.; Zhou,~J.; Rugg,~B.~K.; Krzyaniak,~M.~D.;
  Wasielewski,~M.~R. {CNOT gate operation on a photogenerated molecular
  electron spin-qubit pair}. \emph{J. Chem. Phys.} \textbf{2020}, \emph{152},
  014503\relax
\mciteBstWouldAddEndPuncttrue
\mciteSetBstMidEndSepPunct{\mcitedefaultmidpunct}
{\mcitedefaultendpunct}{\mcitedefaultseppunct}\relax
\EndOfBibitem
\bibitem[Wasielewski \latin{et~al.}(2020)Wasielewski, Forbes, Frank, Kowalski,
  Scholes, Yuen-Zhou, Baldo, Freedman, Goldsmith, Goodson, Kirk, McCusker,
  Ogilvie, Shultz, Stoll, and Whaley]{Wasielewski2020}
Wasielewski,~M.~R.; Forbes,~M. D.~E.; Frank,~N.~L.; Kowalski,~K.;
  Scholes,~G.~D.; Yuen-Zhou,~J.; Baldo,~M.~A.; Freedman,~D.~E.;
  Goldsmith,~R.~H.; Goodson,~T. \latin{et~al.}  {Exploiting chemistry and
  molecular systems for quantum information science}. \emph{Nat. Rev. Chem.}
  \textbf{2020}, \emph{4}, 490--504\relax
\mciteBstWouldAddEndPuncttrue
\mciteSetBstMidEndSepPunct{\mcitedefaultmidpunct}
{\mcitedefaultendpunct}{\mcitedefaultseppunct}\relax
\EndOfBibitem
\bibitem[Rugg \latin{et~al.}(2019)Rugg, Krzyaniak, Phelan, Ratner, Young, and
  Wasielewski]{Rugg2019}
Rugg,~B.~K.; Krzyaniak,~M.~D.; Phelan,~B.~T.; Ratner,~M.~A.; Young,~R.~M.;
  Wasielewski,~M.~R. {Photodriven quantum teleportation of an electron spin
  state in a covalent donor–acceptor–radical system}. \emph{Nat. Chem.}
  \textbf{2019}, \emph{11}, 981--986\relax
\mciteBstWouldAddEndPuncttrue
\mciteSetBstMidEndSepPunct{\mcitedefaultmidpunct}
{\mcitedefaultendpunct}{\mcitedefaultseppunct}\relax
\EndOfBibitem
\bibitem[Basel \latin{et~al.}(2019)Basel, Young, Krzyaniak, Papadopoulos,
  Hetzer, Gao, {La Porte}, Phelan, Clark, Tykwinski, Wasielewski, and
  Guldi]{Basel2019}
Basel,~B.~S.; Young,~R.~M.; Krzyaniak,~M.~D.; Papadopoulos,~I.; Hetzer,~C.;
  Gao,~Y.; {La Porte},~N.~T.; Phelan,~B.~T.; Clark,~T.; Tykwinski,~R.~R.
  \latin{et~al.}  {Influence of the heavy-atom effect on singlet fission: a
  study of platinum-bridged pentacene dimers}. \emph{Chem. Sci.} \textbf{2019},
  \emph{10}, 11130--11140\relax
\mciteBstWouldAddEndPuncttrue
\mciteSetBstMidEndSepPunct{\mcitedefaultmidpunct}
{\mcitedefaultendpunct}{\mcitedefaultseppunct}\relax
\EndOfBibitem
\bibitem[Ieuji \latin{et~al.}(2019)Ieuji, Goushi, and Adachi]{Ieuji2019}
Ieuji,~R.; Goushi,~K.; Adachi,~C. {Triplet–triplet upconversion enhanced by
  spin–orbit coupling in organic light-emitting diodes}. \emph{Nat. Commun.}
  \textbf{2019}, \emph{10}, 5283\relax
\mciteBstWouldAddEndPuncttrue
\mciteSetBstMidEndSepPunct{\mcitedefaultmidpunct}
{\mcitedefaultendpunct}{\mcitedefaultseppunct}\relax
\EndOfBibitem
\end{mcitethebibliography}


%merlin.mbs aipnum4-1.bst 2010-07-25 4.21a (PWD, AO, DPC) hacked
%Control: key (0)
%Control: author (8) initials jnrlst
%Control: editor formatted (1) identically to author
%Control: production of article title (-1) disabled
%Control: page (0) single
%Control: year (1) truncated
%Control: production of eprint (0) enabled
\begin{thebibliography}{14}%
\makeatletter
\providecommand \@ifxundefined [1]{%
 \@ifx{#1\undefined}
}%
\providecommand \@ifnum [1]{%
 \ifnum #1\expandafter \@firstoftwo
 \else \expandafter \@secondoftwo
 \fi
}%
\providecommand \@ifx [1]{%
 \ifx #1\expandafter \@firstoftwo
 \else \expandafter \@secondoftwo
 \fi
}%
\providecommand \natexlab [1]{#1}%
\providecommand \enquote  [1]{``#1''}%
\providecommand \bibnamefont  [1]{#1}%
\providecommand \bibfnamefont [1]{#1}%
\providecommand \citenamefont [1]{#1}%
\providecommand \href@noop [0]{\@secondoftwo}%
\providecommand \href [0]{\begingroup \@sanitize@url \@href}%
\providecommand \@href[1]{\@@startlink{#1}\@@href}%
\providecommand \@@href[1]{\endgroup#1\@@endlink}%
\providecommand \@sanitize@url [0]{\catcode `\\12\catcode `\$12\catcode
  `\&12\catcode `\#12\catcode `\^12\catcode `\_12\catcode `\%12\relax}%
\providecommand \@@startlink[1]{}%
\providecommand \@@endlink[0]{}%
\providecommand \url  [0]{\begingroup\@sanitize@url \@url }%
\providecommand \@url [1]{\endgroup\@href {#1}{\urlprefix }}%
\providecommand \urlprefix  [0]{URL }%
\providecommand \Eprint [0]{\href }%
\providecommand \doibase [0]{http://dx.doi.org/}%
\providecommand \selectlanguage [0]{\@gobble}%
\providecommand \bibinfo  [0]{\@secondoftwo}%
\providecommand \bibfield  [0]{\@secondoftwo}%
\providecommand \translation [1]{[#1]}%
\providecommand \BibitemOpen [0]{}%
\providecommand \bibitemStop [0]{}%
\providecommand \bibitemNoStop [0]{.\EOS\space}%
\providecommand \EOS [0]{\spacefactor3000\relax}%
\providecommand \BibitemShut  [1]{\csname bibitem#1\endcsname}%
\let\auto@bib@innerbib\@empty
%</preamble>
\bibitem [{\citenamefont {Fedorov}\ and\ \citenamefont
  {Gordon}(2000)}]{Fedorov2005}%
  \BibitemOpen
  \bibfield  {author} {\bibinfo {author} {\bibfnamefont {D.~G.}\ \bibnamefont
  {Fedorov}}\ and\ \bibinfo {author} {\bibfnamefont {M.~S.}\ \bibnamefont
  {Gordon}},\ }\href {\doibase 10.1063/1.481136} {\bibfield  {journal}
  {\bibinfo  {journal} {J. Chem. Phys.}\ }\textbf {\bibinfo {volume} {112}},\
  \bibinfo {pages} {5611} (\bibinfo {year} {2000})}\BibitemShut {NoStop}%
\bibitem [{\citenamefont {Fedorov}\ \emph {et~al.}(2003)\citenamefont
  {Fedorov}, \citenamefont {Koseki}, \citenamefont {Schmidt},\ and\
  \citenamefont {Gordon}}]{Fedorov2003}%
  \BibitemOpen
  \bibfield  {author} {\bibinfo {author} {\bibfnamefont {D.~G.}\ \bibnamefont
  {Fedorov}}, \bibinfo {author} {\bibfnamefont {S.}~\bibnamefont {Koseki}},
  \bibinfo {author} {\bibfnamefont {M.~W.}\ \bibnamefont {Schmidt}}, \ and\
  \bibinfo {author} {\bibfnamefont {M.~S.}\ \bibnamefont {Gordon}},\ }\href
  {\doibase 10.1080/0144235032000101743} {\bibfield  {journal} {\bibinfo
  {journal} {Int. Rev. Phys. Chem.}\ }\textbf {\bibinfo {volume} {22}},\
  \bibinfo {pages} {551} (\bibinfo {year} {2003})}\BibitemShut {NoStop}%
\bibitem [{\citenamefont {Fay}, \citenamefont {Lindoy},\ and\ \citenamefont
  {Manolopoulos}(2018)}]{Fay2018}%
  \BibitemOpen
  \bibfield  {author} {\bibinfo {author} {\bibfnamefont {T.~P.}\ \bibnamefont
  {Fay}}, \bibinfo {author} {\bibfnamefont {L.~P.}\ \bibnamefont {Lindoy}}, \
  and\ \bibinfo {author} {\bibfnamefont {D.~E.}\ \bibnamefont {Manolopoulos}},\
  }\href {\doibase 10.1063/1.5041520} {\bibfield  {journal} {\bibinfo
  {journal} {J. Chem. Phys.}\ }\textbf {\bibinfo {volume} {149}},\ \bibinfo
  {pages} {064107} (\bibinfo {year} {2018})},\ \Eprint
  {http://arxiv.org/abs/1808.03211} {arXiv:1808.03211} \BibitemShut {NoStop}%
\bibitem [{\citenamefont {Nakajima}(1958)}]{Nakajima1958}%
  \BibitemOpen
  \bibfield  {author} {\bibinfo {author} {\bibfnamefont {S.}~\bibnamefont
  {Nakajima}},\ }\href {\doibase 10.1143/PTP.20.948} {\bibfield  {journal}
  {\bibinfo  {journal} {Prog. Theor. Phys.}\ }\textbf {\bibinfo {volume}
  {20}},\ \bibinfo {pages} {948} (\bibinfo {year} {1958})}\BibitemShut
  {NoStop}%
\bibitem [{\citenamefont {Zwanzig}(1960)}]{Zwanzig1960}%
  \BibitemOpen
  \bibfield  {author} {\bibinfo {author} {\bibfnamefont {R.}~\bibnamefont
  {Zwanzig}},\ }\href {\doibase 10.1063/1.1731409} {\bibfield  {journal}
  {\bibinfo  {journal} {J. Chem. Phys.}\ }\textbf {\bibinfo {volume} {33}},\
  \bibinfo {pages} {1338} (\bibinfo {year} {1960})}\BibitemShut {NoStop}%
\bibitem [{\citenamefont {Sparpaglione}\ and\ \citenamefont
  {Mukamel}(1988)}]{Sparpaglione1988}%
  \BibitemOpen
  \bibfield  {author} {\bibinfo {author} {\bibfnamefont {M.}~\bibnamefont
  {Sparpaglione}}\ and\ \bibinfo {author} {\bibfnamefont {S.}~\bibnamefont
  {Mukamel}},\ }\href {\doibase 10.1063/1.453922} {\bibfield  {journal}
  {\bibinfo  {journal} {J. Chem. Phys.}\ }\textbf {\bibinfo {volume} {88}},\
  \bibinfo {pages} {3263} (\bibinfo {year} {1988})}\BibitemShut {NoStop}%
\bibitem [{\citenamefont {May}\ and\ \citenamefont {K\"uhn}(2000)}]{May2000}%
  \BibitemOpen
  \bibfield  {author} {\bibinfo {author} {\bibfnamefont {V.}~\bibnamefont
  {May}}\ and\ \bibinfo {author} {\bibfnamefont {O.}~\bibnamefont {K\"uhn}},\
  }\href@noop {} {\emph {\bibinfo {title} {{Charge and Energy Transfer Dynamics
  in Molecular Systems}}}}\ (\bibinfo  {publisher} {Wiley-VCH Verlag GmbH {\&}
  Co. KGaA},\ \bibinfo {year} {2000})\BibitemShut {NoStop}%
\bibitem [{\citenamefont {Hoff}\ \emph {et~al.}(1998)\citenamefont {Hoff},
  \citenamefont {Gast}, \citenamefont {Dzuba}, \citenamefont {Timmel},
  \citenamefont {Fursman},\ and\ \citenamefont {Hore}}]{Hoff1998}%
  \BibitemOpen
  \bibfield  {author} {\bibinfo {author} {\bibfnamefont {A.~J.}\ \bibnamefont
  {Hoff}}, \bibinfo {author} {\bibfnamefont {P.}~\bibnamefont {Gast}}, \bibinfo
  {author} {\bibfnamefont {S.~A.}\ \bibnamefont {Dzuba}}, \bibinfo {author}
  {\bibfnamefont {C.~R.}\ \bibnamefont {Timmel}}, \bibinfo {author}
  {\bibfnamefont {C.~E.}\ \bibnamefont {Fursman}}, \ and\ \bibinfo {author}
  {\bibfnamefont {P.}~\bibnamefont {Hore}},\ }\href {\doibase
  10.1016/S1386-1425(98)00211-X} {\bibfield  {journal} {\bibinfo  {journal}
  {Spectrochim. Acta Part A Mol. Biomol. Spectrosc.}\ }\textbf {\bibinfo
  {volume} {54}},\ \bibinfo {pages} {2283} (\bibinfo {year}
  {1998})}\BibitemShut {NoStop}%
\bibitem [{\citenamefont {Schulten}\ and\ \citenamefont
  {Wolynes}(1978)}]{Schulten1978}%
  \BibitemOpen
  \bibfield  {author} {\bibinfo {author} {\bibfnamefont {K.}~\bibnamefont
  {Schulten}}\ and\ \bibinfo {author} {\bibfnamefont {P.~G.}\ \bibnamefont
  {Wolynes}},\ }\href {\doibase 10.1063/1.436135} {\bibfield  {journal}
  {\bibinfo  {journal} {J. Chem. Phys.}\ }\textbf {\bibinfo {volume} {68}},\
  \bibinfo {pages} {3292} (\bibinfo {year} {1978})}\BibitemShut {NoStop}%
\bibitem [{\citenamefont {Steiner}\ and\ \citenamefont
  {Ulrich}(1989)}]{Steiner1989}%
  \BibitemOpen
  \bibfield  {author} {\bibinfo {author} {\bibfnamefont {U.~E.}\ \bibnamefont
  {Steiner}}\ and\ \bibinfo {author} {\bibfnamefont {T.}~\bibnamefont
  {Ulrich}},\ }\href {\doibase 10.1021/cr00091a003} {\bibfield  {journal}
  {\bibinfo  {journal} {Chem. Rev.}\ }\textbf {\bibinfo {volume} {89}},\
  \bibinfo {pages} {51} (\bibinfo {year} {1989})}\BibitemShut {NoStop}%
\bibitem [{\citenamefont {Rodgers}(2009)}]{Rodgers2009}%
  \BibitemOpen
  \bibfield  {author} {\bibinfo {author} {\bibfnamefont {C.~T.}\ \bibnamefont
  {Rodgers}},\ }\href {\doibase 10.1351/PAC-CON-08-10-18} {\bibfield  {journal}
  {\bibinfo  {journal} {Pure Appl. Chem.}\ }\textbf {\bibinfo {volume} {81}},\
  \bibinfo {pages} {19} (\bibinfo {year} {2009})}\BibitemShut {NoStop}%
\bibitem [{\citenamefont {Mims}\ \emph {et~al.}(2019)\citenamefont {Mims},
  \citenamefont {Schmiedel}, \citenamefont {Holzapfel}, \citenamefont {Lukzen},
  \citenamefont {Lambert},\ and\ \citenamefont {Steiner}}]{Mims2019a}%
  \BibitemOpen
  \bibfield  {author} {\bibinfo {author} {\bibfnamefont {D.}~\bibnamefont
  {Mims}}, \bibinfo {author} {\bibfnamefont {A.}~\bibnamefont {Schmiedel}},
  \bibinfo {author} {\bibfnamefont {M.}~\bibnamefont {Holzapfel}}, \bibinfo
  {author} {\bibfnamefont {N.~N.}\ \bibnamefont {Lukzen}}, \bibinfo {author}
  {\bibfnamefont {C.}~\bibnamefont {Lambert}}, \ and\ \bibinfo {author}
  {\bibfnamefont {U.~E.}\ \bibnamefont {Steiner}},\ }\href {\doibase
  10.1063/1.5131056} {\bibfield  {journal} {\bibinfo  {journal} {J. Chem.
  Phys.}\ }\textbf {\bibinfo {volume} {151}},\ \bibinfo {pages} {244308}
  (\bibinfo {year} {2019})}\BibitemShut {NoStop}%
\bibitem [{\citenamefont {Fay}, \citenamefont {Lindoy},\ and\ \citenamefont
  {Manolopoulos}(2019)}]{Fay2019b}%
  \BibitemOpen
  \bibfield  {author} {\bibinfo {author} {\bibfnamefont {T.~P.}\ \bibnamefont
  {Fay}}, \bibinfo {author} {\bibfnamefont {L.~P.}\ \bibnamefont {Lindoy}}, \
  and\ \bibinfo {author} {\bibfnamefont {D.~E.}\ \bibnamefont {Manolopoulos}},\
  }\href {\doibase 10.1063/1.5125752} {\bibfield  {journal} {\bibinfo
  {journal} {J. Chem. Phys.}\ }\textbf {\bibinfo {volume} {151}},\ \bibinfo
  {pages} {154117} (\bibinfo {year} {2019})}\BibitemShut {NoStop}%
\bibitem [{\citenamefont {Riese}\ \emph {et~al.}(2020)\citenamefont {Riese},
  \citenamefont {Brand}, \citenamefont {Mims}, \citenamefont {Holzapfel},
  \citenamefont {Lukzen}, \citenamefont {Steiner},\ and\ \citenamefont
  {Lambert}}]{Riese2020}%
  \BibitemOpen
  \bibfield  {author} {\bibinfo {author} {\bibfnamefont {S.}~\bibnamefont
  {Riese}}, \bibinfo {author} {\bibfnamefont {J.~S.}\ \bibnamefont {Brand}},
  \bibinfo {author} {\bibfnamefont {D.}~\bibnamefont {Mims}}, \bibinfo {author}
  {\bibfnamefont {M.}~\bibnamefont {Holzapfel}}, \bibinfo {author}
  {\bibfnamefont {N.~N.}\ \bibnamefont {Lukzen}}, \bibinfo {author}
  {\bibfnamefont {U.~E.}\ \bibnamefont {Steiner}}, \ and\ \bibinfo {author}
  {\bibfnamefont {C.}~\bibnamefont {Lambert}},\ }\href {\doibase
  10.1063/5.0013941} {\bibfield  {journal} {\bibinfo  {journal} {J. Chem.
  Phys.}\ }\textbf {\bibinfo {volume} {153}},\ \bibinfo {pages} {054306}
  (\bibinfo {year} {2020})}\BibitemShut {NoStop}%
\end{thebibliography}%

\end{document}

\title{Supporting Information to ``Chirality Induced Spin Coherence in Electron Transfer Reactions''}
\author{Thomas P. Fay}
\email{tom.patrick.fay@gmail.com}
\affiliation{Department of Chemistry, University of Oxford, Physical and Theoretical Chemistry Laboratory, South Parks Road, Oxford, OX1 3QZ, UK}
\maketitle
\tableofcontents

\section{The spin-orbit coupling interaction}\label{soc-app}

In the main paper we simply state that the spin-orbit coupling term in a trunctated diabatic basis containing two charge transfer states is given by Eqs.~(3) and (11). In the one electron case, the electronic state factorises into spatial and spin parts given by $\ket{A,\sigma} = \psi_A(\vb{r}|\vb{Q})\ket{\sigma}$, where $\psi_A(\vb{r}|\vb{Q})$ is the nuclear configuration dependent diabatic state wavefunction. Importantly for bound electronic states these wavefunctions can always be chosen to be real-valued.\cite{Fedorov2005} The full spin orbit coupling Hamiltonian for an $N$ electron system is\cite{Fedorov2005}
\begin{align}
	\op{H}_{\mathrm{SOC}} = \sum_{i=1}^N\sum_{K} \xi_K(r_{iK}) \op{\boldsymbol{\ell}}_{iK} \cdot \op{\vb{s}}_i,
\end{align}
where $r_{iK} = |\vb{r}_i-\vb{Q}_K|$, $\xi_K(r_{iK})$ is a real-valued function of $r_{iK}$ and $\op{\boldsymbol{\ell}}_{iK} = (\vb{r}_i - \vb{Q}_K)\times \op{\vb{p}}_i$ is the orbital angular momentum of electron $i$ about nucleus $K$, $\op{\vb{s}}_i$ is the momentum operator for electron $i$ and $\op{\vb{s}}_i$ is the spin operator for electron $i$. The SOC matrix elements are then given by (noting that for one electron we can drop the electron labels $i$)
\begin{align}
	\begin{split}
	&\mel{A,\sigma}{\op{H}_{\mathrm{SOC}}}{A',\sigma'} = \sum_{\alpha = x,y,z}\mel{\sigma}{\op{s}_\alpha}{\sigma'} 
	\int\dd{\vb{r}} \psi_j(\vb{r}|\vb{Q})^*\sum_{K} \xi_A(r_K) \op{{\ell}}_{K,\alpha} \psi_{A'}(\vb{r}|\vb{Q})
	\end{split}
\end{align}
and therefore we identify the components of the spin orbit coupling vector $i\boldsymbol{\Lambda}_{A,A'}(\vb{Q})$ as
\begin{align}
	i\Lambda_{A,A',\alpha}(\vb{Q}) = \!\!\int\!\!\dd{\vb{r}} \psi_A(\vb{r}|\vb{Q})^*\sum_{K} \xi_K(r_K) \op{{\ell}}_{K,\alpha} \psi_{A'}(\vb{r}|\vb{Q}).
\end{align}
Noting that $\op{\vb{p}} = -i\hbar \boldsymbol{\nabla}_{\vb{r}}$, and that the diabatic wave functions are real-valued, this term must be zero for $A=A'$ and for $A\neq A'$ $\Lambda_{A,A',\alpha}(\vb{Q})$ is real valued with $\Lambda_{A,A',\alpha}(\vb{Q}) = - \Lambda_{A',A,\alpha}(\vb{Q})$. This immediately gives the diabatic basis spin-orbit coupling Hamiltonian given by Eq.~(3) in the main text, when $\psi_A$ is taken to be an orbital localised on \ce{D}, $\psi_\mathrm{D}$, and $\psi_{A'}$ is taken to be an orbital localised on \ce{A}, $\psi_{\mathrm{A}}$, and when we also invoke the Condon approximation i.e. assuming that these matrix elements have only a weak dependence on $\vb{Q}$ for accessible conformations of the molecule.

This argument generalises to the many electron case as follows. For an arbitrary electronic state $\ket{\Psi_A,S=1/2,M_S}$ with total spin quantum number $S=1/2$ with real expansion coefficients in terms of Slater determinants of real-valued molecular orbitals% (i.e. assuming no symmetry related degeneracy of the states)
, it is true that
\begin{align}
	\mel{\Psi_A,1/2,\pm 1/2}{\op{H}_\mathrm{SOC}}{\Psi_{B},1/2,\mp 1/2}  &=   i a_{A,B} \pm b_{A,B} \\
	\mel{\Psi_A,1/2,\pm 1/2}{\op{H}_\mathrm{SOC}}{\Psi_{B},1/2,\pm 1/2} &= \pm i c_{A,B} 
\end{align}
where $a_{A,B}, b_{A,B}$ and $c_{A,B}$ are real valued, and all $0$ if $A=B$ (see Ref.~\citenum{Fedorov2003} for details). For two $S=1/2$ diabatic states $\ket{\Psi_0,1/2,M_S} = \ket{\mathrm{D^{\bullet -}A},1/2,M_S} = \ket{0,1/2,M_S}$ and $\ket{\Psi_1,1/2,M_S}=\ket{\mathrm{DA^{\bullet -}},1/2,M_S} =\ket{1,1/2,M_S}$, the spin orbit coupling Hamiltonian in this truncated basis is therefore
\begin{align}
	\begin{split}
	\op{V}_{\mathrm{SOC}} = & 2i\left(a_{0,1}\op{S}_{x}+b_{0,1}\op{S}_{y} +c_{0,1}\op{S}_{z}\right) \left(\dyad{0}{1} -\dyad{1}{0}\right)
	\end{split}
\end{align}
and we can therefore identify $\boldsymbol{\Lambda} = -2(a_{0,1},b_{0,1},c_{0,1})$. These are implicitly dependent on $\vb{Q}$ but again we can invoke the Condon approximation to ignore this $\vb{Q}$ dependence.

In order to understand the two unpaired electron radical pair case, we will first consider a two electron system, in which orbitals $\psi_{A}(\vb{r}|\vb{Q})$ are occupied, and the total electronic states are denoted $\ket{A,B,\sigma_1,\sigma_2}$ which are the antisymmetrised states in which orbital $A$ is occupied with spin $\sigma_1$ and orbital $B$ is occupied with spin $\sigma_2$,
\begin{align}
	\begin{split}
		\ket{A,B,\sigma_1,\sigma_2} &= \frac{1}{\sqrt{2}}\bigg(\psi_{A}(\vb{r}_1)\psi_{B}(\vb{r}_2)\ket{\sigma_1}_1\ket{\sigma_2}_2 -\psi_{B}(\vb{r}_1)\psi_{A}(\vb{r}_2)\ket{\sigma_2}_1\ket{\sigma_1}_2\bigg),
	\end{split}
\end{align}
where $\ket{\sigma}_j$ is a spin state of electron $j$, and $\vb{r}_j$ refer to the coordinates of electron $j$, and we have dropped the explicit dependence on $\vb{Q}$ for brevity. In our case we consider the coupling between a state $\ket{0}$ where one electron is the donor HOMO $\psi_{\mathrm{D}}$, and one in the donor LUMO $\psi_{\mathrm{D^*}}$ (which corresponds to the photoexcited precursor state), and a state $\ket{1}$ where one electron is in the donor HOMO $\psi_{\mathrm{D}}$ and one in the acceptor LUMO $\psi_{\mathrm{A}}$ (which corresponds to the charge separated radical pair state). Because the spin orbit coupling matrix element between identical orbitals is zero, the only non-zero SOC matrix elements between these states can be written as
\begin{align}
	\begin{split}
	&\mel{\mathrm{D}^*,\mathrm{D},\sigma_1,\sigma_2}{\op{H}_{\mathrm{soc}}}{\mathrm{A},\mathrm{D},\sigma_{1}',\sigma_2'} \!=\!\! \sum_{K,\alpha } \braket{\sigma_2}{{\sigma'_2}}\\
	&\times\mel{\sigma_1}{\op{s}_{\alpha}}{{\sigma_{1}'}}\int\dd{\vb{r}}\psi_{\mathrm{D}^*}(\vb{r})^* \xi_K(r_K)\op{{\ell}}_{K,\alpha}\psi_{\mathrm{A}}(\vb{r}).
	\end{split}
\end{align}
As before the spin-orbit coupling integral between $\psi_{\mathrm{D}^*}$ and $\psi_{\mathrm{A}}$ is purely imaginary valued. We define the effective electron spin operator $\op{\vb{S}}_1$ on the truncated set of states $\ket{A,B,\sigma_1,\sigma_2}$ in terms of the matrix elements of an $S=1/2$ spin operator $\op{\vb{s}}$ i.e. $\mel{A,B,\sigma_1,\sigma_2}{\op{\vb{S}}_1}{A',B',\sigma'_1,\sigma'_2} = \mel{\sigma_1}{\op{\vb{s}}}{\sigma'_1}\delta_{A,A'}\delta_{B,B'}\delta_{\sigma_2,\sigma'_2}$. We can now construct the spin orbit coupling operator on the trunctated set of states $\ket{0} = \ket{0,\sing} = \frac{1}{\sqrt{2}}(\ket{\mathrm{D}^*,\mathrm{D},\alpha,\beta} - \ket{\mathrm{D}^*,\mathrm{D},\beta,\alpha})$ and $\ket{1,\sigma_1,\sigma_2} = \ket{\mathrm{A},\mathrm{D},\sigma_1,\sigma_2}$, exactly as is given in Eq.~(11) in the main text.

Again we can generalise this to a many electron system by noting that for arbitrary elecronic states $\ket{\Psi_A,S,M_S}$ which have total electronic spin quantum number $S=0$ or $S=1$ and total electron spin projection quantum number $M_S$, provided they can be expanded as a real linear combination of Slater determinants of real valued molecular orbitals, then the spin orbit coupling matrix elements have the following properties\cite{Fedorov2003}
\begin{align}
	\mel{\Psi_A,0,0}{\op{H}_{\mathrm{soc}}}{\Psi_{B},1,\pm 1} &= \mp i a_{A,B} + b_{A,B} \\
	\mel{\Psi_A,0,0}{\op{H}_{\mathrm{soc}}}{\Psi_{B},1,0} &= i c_{A,B},
\end{align}
where again the constants are all real-valued. Now considering just a subset of diabatic states $\ket{\Psi_{0},0,0} =\ket{\mathrm{D^*A},0,0}=\ket{0,0,0} $, and $\ket{\Psi_{1},S,M_S} = \ket{\mathrm{D^{\bullet+}A^{\bullet -}},S,M_S}= \ket{1,S,M_S}$ which correspond to the precursor state and the set of $S=0$ and $S=1$ radical pair states, and ignoring the spin-orbit coupling between radical pair states (which vanishes for large radical separations), the spin orbit coupling operator can be written as in Eq.~(11) of the main text with the above definition of the effective electron spin operator $\op{\vb{S}}_1$ with
\begin{align}
	\boldsymbol{\Lambda} = (2\sqrt{2}a_{0,1},2\sqrt{2}b_{0,1},2c_{0,1}).
\end{align}

\section{Second order master equation theory}

In this appendix we briefly summarise the master equation used to obtain the spin density operator master equations in the main text. Further details can be found in Ref.~\citenum{Fay2018}. Our starting point is the Nakajima-Zwanzig equation\cite{Nakajima1958,Zwanzig1960} for a projected density operator $\pP\op{\rho}(t)$, which contains the spin density operators,\cite{Fay2018}
\begin{align}
	\dv{t}\pP \op{\rho}(t) = \pP \pL \pP\op{\rho}(t) + \int_0^t\pK(t-\tau)\pP\op{\rho}(\tau)\dd{\tau},
\end{align}
where $\pL = -(i/\hbar)[\op{H},\ \cdot \ ]$ is the total Liouvillian for the system and $\pK(t)$ is a memory kernel which contains the effects of parts of the system removed by projecting the density operator with $\pP$. Analogous to previous work, we define $\pP = \sum_{j} \op{\rho}_{j\nuc}\Tr_\nuc[\op{\Pi}_j\ \cdot  \ \op{\Pi}_j]$, and we expand the kernel $\pK(t)$ to second order in the coupling $\Gamma$. %Furthermore we make the Markovian approximation to the term involving the kernel, and we ignore any spin terms $\op{H}_{j\sys}$ in evaluating $\pK(t)$, both of which are valid approximations if the nuclear timescales are much faster the the timescales on which the spin density operators evolve. [A summary of the theory is given in the SI, and we refer the reader to Ref.~\citenum{Fay2018} for more details.]

The approximate treatment of the spin-orbit and diabatic coupling is based on approximating the kernel in the Nakajima-Zwanzig equation, which is given by
\begin{align}
	\pK(t) = \pP \pL e^{(1-\pP)\pL t}(1-\pP) \pL \pP.
\end{align}
Expanding this up to second order in the spin-orbit coupling and diabatic coupling, and ignoring the spin terms $\op{H}_{j\sys}$ in the kernel gives the following approximation to $\pK(t)$
\begin{align}
	\pK(t)\pP \op{\rho} \approx -\frac{1}{\hbar^2}\sum_{j} \op{\rho}_{j\nuc}\Tr_\nuc\left[\op{\Pi}_j [\op{V},[\op{V}^\nuc(t),\pP\op{\rho}]]\op{\Pi}_j\right]
\end{align}
where $\op{V} = \op{V}_\mathrm{DC} + \op{V}_{\mathrm{SOC}}$ and $\op{V}^\nuc (t) = e^{-i\op{H}_\nuc t / \hbar} \op{V} e^{+i\op{H}_\nuc t / \hbar}$, where $\op{H}_\nuc = \op{H}_{0\nuc}\op{\Pi}_0 + \op{H}_{1\nuc}\op{\Pi}_1$. We also invoke the Markovian approximation to the time-convolution term, in which we assume the decay time of the kernel is much faster than the characteristic timescales of the spin and population dynamics, and therefore we can approximate the time convolution in the Nakajima-Zwanzig equation as\cite{Sparpaglione1988}
\begin{align}
	\int_0^t \pK(t-\tau) \pP\op{\rho}(\tau) \dd{\tau} \approx \int_0^\infty  \pK(\tau) \dd{\tau}\pP\op{\rho}(t).
\end{align}
Using this approximation we recover the incoherent kinetic description of the electron transfer processes.

By using these approximations and expanding the double commutator in the second order kernel, one arrives at the master equations given above, with $k_\mathrm{f}$ given by the standard Fermi's Golden rule rate expression,\cite{May2000}
\begin{align}
	k_\mathrm{f} = \frac{2\Gamma^2}{\hbar^2} \Re \int_0^\infty \Tr_\nuc[ e^{+i \op{H}_{0\nuc} t/\hbar}e^{-i \op{H}_{1\nuc} t/\hbar}\op{\rho}_{0 \nuc}]\dd{t}
\end{align}
and similarly $k_\mathrm{b}$ is given by
\begin{align}
	k_\mathrm{b} = \frac{2\Gamma^2}{\hbar^2} \Re \int_0^\infty \Tr_\nuc[ e^{+i \op{H}_{1\nuc} t/\hbar}e^{-i \op{H}_{0\nuc} t/\hbar}\op{\rho}_{1 \nuc}]\dd{t}
\end{align}
and the $\delta J$ term in the radical pair master equation is
\begin{align}
	\delta J = \frac{\Gamma^2}{2\hbar}\Im \int_0^\infty \Tr_\nuc[ e^{+i \op{H}_{0\nuc} t/\hbar}e^{-i \op{H}_{1\nuc} t/\hbar}\op{\rho}_{0 \nuc}]\dd{t}.
\end{align}
The derivation of these expressions is described in detail in Ref.~\citenum{Fay2018}. We note that going to higher orders in $\op{\Gamma}$, the $S=1/2$ system master equation is unchanged, but at fourth order and above in $\Gamma$ in the radical pair master equation a decoherence term of the following form will also appear in the master equation,
\begin{align}
	k_\mathrm{D} \left(\op{U}\op{P}_\sing \op{U}^\dag \op{\sigma}_{1\sys}(t) \op{U}\op{P}_\sing \op{U}^\dag -\frac{1}{2}\left\{\op{U}\op{P}_\sing \op{U}^\dag,\op{\sigma}_{1\sys}(t)\right\}\right).
\end{align}
Also the expressions for the master equation parameters $k_\mathrm{f}$, $k_\mathrm{b}$ and $\delta J$ must be corrected at higher orders in $\Gamma$, but the general form of the master equation remains unchanged other than the additional decoherence term.

We note finally that in all of the above, we have used the Condon approximation, in which we assume $\Delta$ and $\boldsymbol{\Lambda}$ are independent of $\vb{Q}$. However in reality these terms will be nuclear configuration dependent and in general they will have a different dependence on $\vb{Q}$. Accounting for this will introduce terms which lead to spin decoherence on electron transport, i.e. terms which only transfer populations and not spin coherences. 

\section{OOP-ESEEM Signals for chiral radical pairs}

Here we outline the derivation of the OOP-ESEEM signal for a chiral radical pair formed by SOCT. We consider the final time dependent $x$ channel FID signals in the rotating frame $f_x(t) = \Tr_\sys[\op{S}_{x}\op{\sigma}_\sys(t)]$. Here $\sigma_{\sys}(t)$ is the radical pair spin density operator (denoted $\op{\sigma}_{1\sys}(t)$ above) after the OOP-ESEEM pulse sequence, and $\op{S}_\alpha = \op{S}_{1\alpha}+\op{S}_{2\alpha}$. For simplicity in finding $f_x(t)$, we will invoke the high field (secular) approximation for the radical pair spin Hamiltonian in the rotating frame,\cite{Hoff1998}
\begin{align}\label{hf-spin-ham-eq}
	\begin{split}
		\frac{\op{H}_\sys}{\hbar} &= \Omega_1\op{S}_{1,z} + \Omega_2\op{S}_{1,z} - 2J \op{\vb{S}}_1\!\cdot \!\op{\vb{S}}_2 
		+ \frac{1}{2}d\left(\op{S}_z^2-\frac{1}{3}\op{S}^2\right) +\sum_{i=1}^2\sum_{k=1}^{N_{i}}a_{i,k}\op{\vb{I}}_{i,k}\cdot \op{\vb{S}}_i.
	\end{split}
\end{align}
From right to left these terms describe the Zeeman interactions of two electron spins (relative to the rotating frame frequency), the scalar electron spin coupling, the dipolar coupling (with $d = D(3\cos^2(\xi)-1)$, where $D$ is the dipolar coupling constant), and the hyperfine coupling terms. 

For simplicity we ignore chiral spin-orbit effects on evolution of the radical pair spin density, for example terms of the form $2\delta J\op{U}\op{P}_\sing\op{U}^\dagger$, and as a further simplification we will ignore any decay processes of the radical pair. This means we will assume the radical pair spin density operator evolves under the equation,
\begin{align}
	\dv{t}\op{\sigma}_\sys(t) = -\frac{i}{\hbar} \left[\op{H}_\sys,\op{\sigma}_\sys(t)\right] = \pL_\sys \op{\sigma}_\sys(t).
\end{align}
The action of a perfect (instantaneous) microwave pulse on the spin density operator by an angle $\varphi$, about an axis $\alpha$ is
%\begin{align}
$	\pazocal{U}_\alpha(\varphi) \op{\sigma}_\sys = e^{-i\varphi\op{S}_{\alpha}} \op{\sigma}_\sys e^{+i\varphi\op{S}_{\alpha}}.$ 
%\end{align}
Finally we assume the nuclear spins are in the infinite temperature thermal equilibrium state, i.e. a completely mixed state, at $t=0$, therefore the initial spin density operator of a chiral radical pair is $\op{\sigma}_\sys(0) = (1/Z)\op{U}\op{P}_\sing\op{U}^\dag $, where $Z$ is the dimensionality of the nuclear spin Hilbert space. The spin density operator in the rotating frame at a time $t$ of the echo, $\op{\sigma}_{\sys}(t)$, is then found as 
\begin{align}
	\op{\sigma}_{\sys}(t) = e^{\pL_\sys t} e^{\pL_\sys\tau}\pazocal{U}_x(\pi) e^{\pL_\sys\tau}\pazocal{U}_x(\pi/4) \op{\sigma}_\sys(0).
\end{align}

We will now use this to calculate the OOP-ESEEM signal when the radical pair is oriented such that $\vb{n}$ is parallel to the lab $z$ axis. We note that the Hamiltonian commutes with $\op{I}_{i,k,z}$, and therefore its eigenstates can be written as $\ket{n,\vb{M}}$, where $\vb{M}$ denotes the set of $z$ projection quantum numbers for the nuclear spins. For simplicity, we make the weak-coupling approximation here, $|\Omega_1 - \Omega_2|\gg |J-d/6| $, and then using standard expressions for the radical pair eigenstates,\cite{Hoff1998} one finds the total $x$ channel OOP-ESEEM signal to be 
%\begin{align}
%\begin{split}
%	&f_x(t) = \frac{1}{Z}\sum_{\vb{M}}\sin \left({\theta}\right) \cos \left(\frac{1}{2} t \left(\omega_{1,\vb{M}}+\omega_{2,\vb{M}}\right)\right) \\
%	&\times\sin \left(\frac{1}{3} (d-3 J) (t+2 \tau )\right) \\ 
%	&\times\bigg(\sqrt{2} \cos \left({\theta}\right) \sin \left(\frac{1}{2} t \left(\omega_{1,\vb{M}}-\omega_{2,\vb{M}}\right)\right)\\
%	&-\sin \left({\theta}\right) \cos \left(\frac{1}{2} t \left(\omega_{1,\vb{M}}-\omega_{2,\vb{M}}\right)\right)\bigg)
%\end{split}
%\end{align}
\begin{align}
	\begin{split}
		&f_x(t) = \frac{\sin \left({\theta }\right)}{Z}\sum_{\vb{M}} \cos \left( \bar{\omega}_{\vb{M}}t\right) \sin \left(\left(\frac{d}{3} -J\right)(t+2 \tau )\right) \\ 
		&\times\bigg(\sqrt{2} \cos \left({\theta}\right) \sin \left( \Delta\omega_{\vb{M}} t \right)-\sin \left({\theta}\right) \cos \left( \Delta\omega_{\vb{M}} t\right)\bigg)
	\end{split}
\end{align}
where $\bar{\omega}_{\vb{M}} = (\omega_{1,\vb{M}}+\omega_{2,\vb{M}})/2$, $\Delta{\omega}_{\vb{M}} = (\omega_{1,\vb{M}}-\omega_{2,\vb{M}})/2$ $\omega_{i,\vb{M}} = \Omega_i + \sum_{k=1}^{N_i} a_{i,k}M_{i,k}$. In the semiclassical limit, where the number of hyperfine coupled nuclear spins is larger, we can replace $\sum_{k=1}^{N_i} a_{i,k}M_{i,k} \to h_{i,z}$, and $\frac{1}{Z}\sum_{\vb{M}} \to \prod_{i=1}^2\int_{-\infty}^{\infty}\dd{h_{i,z}}P_{i,z}(h_{i,z})$, where the probability distribution is given by\cite{Schulten1978}
\begin{align}
	P_{i,z}(h_{i,z}) = \frac{\tau_{i}}{\sqrt{2\pi}} \exp(-\frac{1}{2}\left({h_{i,z}}{\tau_{i}}\right)^2)
\end{align}
where $\tau_{i}^{-2} = \frac{1}{3}\sum_{k=1}^{N_i} a_{i,k}^2 I_{i,k}(I_{i,k}+1)$. This gives the expression for $f_x(t)$ in the weak coupling limit given in the main text.

%Here we present additional analytical expressions for OOP-ESEEM observables. 

Beyond the weak coupling limit, but still within the high field approximation, we can obtain analytical expressions for the $x$ channel FID signal,
\begin{align}
	\begin{split}
		&f_x(t) = \frac{1}{Z}\sum_{\vb{M}}\frac{1}{2} \cos(t \bar{\omega}_{\vb{M}}) \bigg(2 \sin ^2(\theta) \cos(\Delta_{\vb{M}}  \tau ) \sin ((J-d/3) (t+2 \tau )) \cos (\Delta_{\vb{M}}  (t+\tau ))\\
		&-2 \sin ^2(\theta) \cos (4 \phi_{\vb{M}} ) \sin (\Delta_{\vb{M}}  \tau ) \sin ((J-d/3) (t+2 \tau )) \sin (\Delta_{\vb{M}}  (t+\tau ))\\
		&-2 \sin ^2(\theta) \cos (2 \phi_{\vb{M}} ) \cos ((J-d/3) (t+2 \tau )) \sin (\Delta_{\vb{M}}  (t+2 \tau ))\\
		&-\sqrt{2} \sin (\theta ) \big(\sin (2 \phi_{\vb{M}} ) \sin (\Delta_{\vb{M}}  t) \sin ((J-d/3) (t+2 \tau ))\\
		&+\sin (4 \phi_{\vb{M}} ) \sin (\Delta_{\vb{M}}  \tau ) \cos ((J-d/3) (t+2 \tau )) \sin (\Delta_{\vb{M}}  (t+\tau ))\big)\bigg)
	\end{split}
\end{align}
where $\bar{\omega}_{\vb{M}} = (\omega_{1,\vb{M}}+\omega_{2,\vb{M}})/2$, $\Delta_{\vb{M}} = \sqrt{\Delta{\omega}_{\vb{M}}^2+(J+d/6)^2}$, $\tan(2\phi_{\vb{M}}) = \Delta\omega_{\vb{M}} / (J+d/6)$, $\Delta{\omega}_{\vb{M}} = (\omega_{1,\vb{M}}-\omega_{2,\vb{M}})/2$, and $\omega_{i,\vb{M}} = \Omega_i + \sum_{k=1}^{N_i} a_{i,k}M_{i,k}$. By making the same semiclassical replacement as above, we can numerically evaluate the semiclassical FID signals as a function of $t$. We can then numerically integrate these for a range of $\tau$ values to yield $F_x(\tau)$.

In the weak coupling, semiclassical limit the integrated FID signal can be found analytically as a function of $\tau$ by integrating Eq.~(18) as $F_x(\tau) = \int_0^\infty f_x(t)\dd{t}$,
\begin{widetext}
\begin{align}
	\begin{split}
		&F_x(\tau) = \frac{\sqrt{\pi}}{8}\bigg[\tau_1 \left(e^{-\tau_1^2(J-d/3+\Omega_1)^2/2}-e^{-\tau_1^2(J-d/3-\Omega_1)^2/2}\right)\\
		&+\tau_2 \left(e^{-\tau_2^2(J-d/3-\Omega_2)^2/2}-e^{-\tau_2^2(J-d/3+\Omega_2)^2/2}\right)\bigg]\cos(2(J-d/3)\tau)\sin(\theta) \\
		&+\frac{1}{4}\bigg[\tau_1 \left(D_+(\tau_1(J-d/3-\Omega_1)/\sqrt{2})-D_+(\tau_1(J-d/3+\Omega_1)/\sqrt{2})\right)\\
		&+\tau_2 \left(D_+(\tau_2(J-d/3+\Omega_2)/\sqrt{2})-D_+(\tau_2(J-d/3-\Omega_2)/\sqrt{2})\right)\bigg] \sin(2(J-d/3)\tau)\sin(\theta) \\
		&+\frac{1}{4}\sqrt{\frac{\pi}{2}}\bigg[\tau_1 \left(e^{-\tau_1^2(J-d/3+\Omega_1)^2/2}+e^{-\tau_1^2(J-d/3-\Omega_1)^2/2}\right)\\
		&+\tau_2 \left(e^{-\tau_2^2(J-d/3-\Omega_2)^2/2}+e^{-\tau_2^2(J-d/3+\Omega_2)^2/2}\right)\bigg]\sin(2(J-d/3)\tau)\sin[2]({\theta}) \\
		&+\frac{1}{2\sqrt{2}}\bigg[\tau_1 \left(D_+(\tau_1(J-d/3-\Omega_1)/\sqrt{2})+D_+(\tau_1(J-d/3+\Omega_1)/\sqrt{2})\right)\\
		&+\tau_2 \left(D_+(\tau_2(J-d/3+\Omega_2)/\sqrt{2})+D_+(\tau_2(J-d/3-\Omega_2)/\sqrt{2})\right)\bigg]\cos(2(J-d/3)\tau)\sin[2]({\theta}) 
	\end{split}
\end{align}
\end{widetext}
where $D_+(x) = e^{-x^2}\int_0^x e^{\lambda^2}\dd{\lambda}$ is the Dawson function.

\section{Magnetic field effects}

Chirality induced spin coherence in oriented radical pairs could also manifest in magnetic field effect experiments, in which quantum yields of spin selective radical pair reactions, or the lifetime of radical pair state, are measured as the applied magnetic field strength is varied, for example by transient absorption spectroscopy.\cite{Steiner1989,Rodgers2009}

As a simple illustrative example, as above, we consider a radical pair oriented such that the spin orbit coupled charge transport vector $\boldsymbol{\Lambda}$ is aligned parallel to the applied magnetic field. We also take the high field limit where the $\Delta g$ mechanism dominates the singlet-triplet interconversion. Treating the spin selective recombination process with the standard Haberkorn reaction term in the master equation for $\op{\sigma}_\sys(t)$, i.e. $- \left\{(k_\sing / 2)\op{P}_\sing + (k_\trip/2)\op{P}_\trip, \op{\sigma}_\sys(t)\right\}$, and assuming that $|\Omega_1-\Omega_2|\gg |J|, |D|, |k_\sing  - k_\trip|, 1/\tau_{i}$, we can use Eq.~\eqref{hf-spin-ham-eq}, to find the triplet quantum yield for a chiral radical pair, $\Phi_\trip = k_\trip\int_0^\infty \Tr_\sys[\op{P}_\trip\op{\sigma}_{\sys}(t)]\dd{t}$ in the high field limit,
\begin{align}
	\Phi_\trip = k_\trip\frac{4 \Delta \Omega ^2-\bar{k}^2 \cos (2\theta )+\bar{k}^2-2 \Delta \Omega  \bar{k} \sin (2\theta )}{2 \bar{k}\left(\bar{k}^2+4 \Delta \Omega ^2 \right)},
\end{align}
where $\Delta\Omega = \Omega_1 - \Omega_2$ and $\bar{k}= (k_\sing +k_\trip)/2$. The term proportional to $\sin(2\theta)$ leads to a chirality dependent triplet quantum yield, and from this the difference in the triplet yields between two enantiomers is maximised when $|\Delta\Omega| = \bar{k}/2$.

Of course, the quantitative interpretation of any real experiment would require more detailed quantum mechanical modelling, accounting fully for hyperfine interactions and spin relaxation,\cite{Mims2019a,Fay2019b,Riese2020} but this simple result demonstrates that chirality induced spin coherence could in principle lead to chirality dependent magnetic field effects on the reactions of oriented radical pairs.

\bibliography{chiral-radical-pair-soct.bib}